\documentclass[aps,twocolumn,showpacs,preprintnumbers,floatfix]{revtex4}
\usepackage{graphicx}
\usepackage{epsfig}
\usepackage{dcolumn}
\usepackage{bm}
\begin{document}

\title{Glueballs at Finite Temperature in $SU(3)$ Yang-Mills Theory}
\author{
          Xiang-Fei Meng${}^{ab}$,
          Gang Li${}^{cd}$,
          Yuan-Jiang Zhang${}^{cd}$,
          Ying Chen${}^{cd}$,
          Chuan Liu${}^{e}$,
          Yu-Bin Liu${}^{a}$,
          Jian-Ping Ma${}^{f}$, and
          Jian-Bo Zhang${}^{g}$\\
         (CLQCD Collaboration)}
\affiliation {
          ${}^a$School of Physics, Nankai University, Tianjin 300071, People¡¯s Republic of China\\
          ${}^b$National Supercomputing Center, Tianjin 300457, People¡¯s Republic of China\\
          ${}^c$Institute of High Energy Physics, Chinese Academy of Sciences, Beijing 100049, People¡¯s Republic of China \\
          ${}^d$Theoretical Center for Science Facilities, Chinese Academy of Sciences, Beijing 100049, People¡¯s Republic of China\\
          ${}^e$School of Physics, Peking University, Beijing 100871, People¡¯s Republic of China\\
          ${}^f$Institute of Theoretical Physics, Chinese Academy of Sciences, Beijing 100080, People¡¯s Republic of China\\
          ${}^g$Department of Physics, Zhejiang University, Hangzhou, Zhejiang 310027, People¡¯s Republic of China
}

\begin{abstract}
{Thermal properties of glueballs in $SU(3)$ Yang-Mills theory are
investigated in a large temperature range from $0.3T_c$ to $1.9T_c$
on anisotropic lattices. The glueball operators are optimized for
the projection of the ground states by the variational method with a
smearing scheme. Their thermal correlators are calculated in all 20
symmetry channels. It is found in all channels that the pole masses
$M_G$ of glueballs remain almost constant when the temperature is
approaching the critical temperature $T_c$ from below, and start to
reduce gradually with the temperature going above $T_c$. The
correlators in the $0^{++}$, $0^{-+}$, and $2^{++}$ channels are
also analyzed based on the Breit-Wigner $\emph{Ansatz}$ by assuming
a thermal width $\Gamma$ to the pole mass $\omega_0$ of each thermal
glueball ground state. While the values of $\omega_0$ are
insensitive to $T$ in the whole temperature range, the thermal
widths $\Gamma$ exhibit distinct behaviors at temperatures below and
above $T_c$. The widths are very small (approximately few percent of
$\omega_0$ or even smaller) when $T<T_c$, but grow abruptly when
$T>T_c$ and reach values of roughly $\Gamma\sim \omega_0/2$ at
$T\approx 1.9T_c$. }
\end{abstract}
 \maketitle

\section{Introduction}

The past two or three decades witnessed intensive and extensive
studies on the phase transition of quantum
chromodynamics(QCD)~\cite{cpod2007}, which is believed to be the
fundamental theory of strong interaction. Based on the two
characteristics of QCD, namely the conjectured color confinement at
low energies and the asymptotic freedom of gluons and quarks at high
energies, QCD at finite temperature is usually described by two
extreme pictures. One is with the weakly interacting meson gas in
the low temperature regime and another is with perturbative quark
gluon plasma (QGP) in the high temperature regime. The two regimes
are bridged by a deconfinement phase transition (or crossover). The
study of the equation of state shows that the perturbative picture
of QGP can only be achieved at very high temperatures $T\ge 2T_c$.
In other words, the dynamical degrees of freedom up to the
temperature of a few times of $T_c$ are not just the quasifree
gluons and quarks~\cite{EOS}. Some other theoretical studies also
support this scenario and conjecture that in the intermediate
temperature range above $T_c$ there may exist different types of
excitations corresponding to different distance
scales~\cite{soft_modes, quark_number} rendering the thermal states
much more complicated. Apart from the quasifree quarks and gluons at
the small distance scale, the large scale excitations can be
effective low-energy modes in the mesonic channels as a result of
the strongly interacting partons~\cite{gupta}. The properties of the
interaction among quarks and gluons at low and high temperatures can
be studied with thermal correlators.
\par
There have been many works on the correlators of charmonia at finite
temperature. Phenomenological studies predicted the binding between
quarks is reduced to dissolve $J/\psi$ at temperatures close to
$T_c$ and proposed the suppression of charmonia as a signal of
QGP~\cite{plb178, prl57}. For example, potential model studies show
that excited states like $\psi^{'}$ and $\chi_{c}$ are dissociated
at $T_{c}$, while the ground state charmonia $J/\Psi$ and $\eta_{c}$
survive up to
$T=1.1T_{c}$~\cite{zpc37,plb514,prd64,prd70,prc72,epjc43}. However,
it is unclear whether the potential model works well at finite
temperatures~\cite{epjc43b}. In contrast, many recent numerical
studies indicate that $J/\Psi$ and $\eta_{c}$ might still survive
above $1.5T_{c}$~\cite{prl92,prd69,prd74,jhw2005,npa783}. Of course,
it is possible  that the $\bar{c}c$ states observed in lattice QCD
are just scattering states. A further lattice study on spatial
boundary-condition dependence of the energy of low-lying $\bar{c}c$
system concludes that they are spatially localized (quasi)bound
states in the temperature region of $1.11\sim 2.07
T_c$~\cite{ptps174}. Obviously, the results of numerical lattice QCD
studies are coincident to the picture of the QCD transition in the
intermediate temperature regime.

Until now most of the lattice studies on hadronic correlators are in
the quenched approximation. Because of the lack of dynamical quarks
in quenched QCD the binding of quark-antiquark systems must be
totally attributed to the nonperturbative properties of gluons,
which are the unique dynamical degree of freedom in the theory.
Since glueballs are the bound states of gluons, a natural question
is how glueballs respond to the varying temperatures.  At low
temperature $T\sim 0$, the existence of quenched glueballs have been
verified by extensive lattice numerical studies, and their spectrum
are also established quite
well~\cite{prd56,prd60,prd73,npb221,plb309,npb314,prl75,outp}. An
investigation of the evolution of glueballs versus the increasing
temperature is important to understand the QCD
transition~\cite{ptn58,npa637} and the hadronization of quark-gluon
plasma~\cite{prd75}. From the point of view of QCD sum rules,
glueball masses are closely related to the gluon condensate. Lattice
studies~\cite{lap583} and model calculations~\cite{pan70} indicate
that the gluon condensate keeps almost constant below $T_c$ and
reduces gradually with the increasing temperature above $T_c$. Based
on this picture, it is expected intuitively that glueball masses
should show a similar behavior also until they melt into
gluons~\cite{qcd20}. In fact, there has already been a lattice study
on the scalar and tensor glueball properties at finite
temperature~\cite{prd66}. In contrast to the expectation and the
finite $T$ behavior of charmonium spectrum, it is interestingly
observed that the pole-mass reduction starts even below $T_c$
($m_{G}$(T $\sim$ $T_{c}$) $\simeq$ 0.8$m_{G}$(T $\sim$ 0)). It is
known that the spatial symmetry group on the lattice is the
24-element cubic point group $O$, whose irreducible representations
are $R=$ $A_1$, $A_2$, $E$, $T_1$, and $T_2$. Along with the parity
$P$ and charge conjugate transformation $C$, all the possible
quantum numbers that glueballs can catch are $R^{PC}$ with
$PC=++,-+,+-,$, and $++$, which add up to 20 symmetry channels.
Motivated by the different temperature behaviors of $\bar{c}c$
systems with different quantum numbers, we would like to investigate
the temperature dependence of glueballs in this paper.

Our numerical study in this work is carried out on anisotropic
lattices with much finer lattice in the temporal direction than in
spatial ones. In order to explore the temperature evolution of
glueball spectrum, the temperature range studied here  extends from
$0.3T_c$ to $1.9T_c$, which is realized by varying the temporal
extension of the lattice. Using anisotropic lattices, the lattice
parameters are carefully determined so that there are enough time
slices for a reliable data analysis even at the highest temperature.
In the present study, we are only interested in the ground state in
each symmetry channel $R^{PC}$. For the study optimized glueball
operators that couple mostly to the ground states are desired.
Practically, these optimized operators are built up by the
combination of smearing schemes and the variational
method~\cite{prd56,prd60,prd73}. In the data processing, the
correlators of these optimized operators are analyzed through two
approaches. First, the thermal masses $M_G$ of glueballs are
extracted in all the channels and all over the temperature range by
fitting the correlators with a single-cosh function form, as is done
in the standard hadron mass measurements. Thus the $T_{-}evolution$
of the thermal glueball spectrums is obtained. Secondly, with
respect that the finite temperature effects may result in mass
shifts and thermal widths of glueballs, we also analyze the
correlators in $A_1^{++}$, $A_1^{-+}$, $E^{++}$, and $T_2^{++}$
channels with the Breit-Wigner $\emph{Ansatz}$ which assumes these
glueball thermal widths, say, change $M_G$ into $\omega_0-i\Gamma$
in the spectral function (see below). As a result, the temperature
dependence of $\omega_0$ and $\Gamma$ can shed some light on the
scenario of the QCD transition.

This paper is organized as follows. In Sec. II, a description of the
determination of working parameters, such as the critical
temperature $T_c$, temperature range,  and lattice spacing $a_s$, as
well as a brief introduction to the variational method is given. In
Sec. III, after a discussion of its feasibility, the results of the
single-cosh fit to the thermal correlators are described in details.
The procedure of the Breit-Wigner fit is also given in this section.
Section IV gives the conclusion and some further discussions.

\section{Numerical Details}
For heavy particles such as charmonia and glueballs, the
implementation of anisotropic lattices is found to be very efficient
in the previous numerical lattice QCD studies both at low and finite
temperatures. On the other hand, the Symanzik improvement and
tadpole improvement schemes of the gauge action are verified to have
better continuum extrapolation behaviors for many physical
quantities. In other words, the finite lattice spacing artifacts are
substantially reduced by these improvements. With these facts, we
adopt the following improved gauge action which has been extensively
used in the study of glueballs~\cite{prd56,prd60,prd73},
\begin{eqnarray}
&& S_{IA}={\beta}{\{\frac{5}{3}\frac{\Omega_{sp}}{{\xi}u_{s}^{4}}
+\frac{4}{3}\frac{\xi\Omega_{tp}}{u_{t}^{2}u_{s}^{2}}
 -\frac{1}{12}\frac{\Omega_{sr}}{{\xi}u_{s}^{6}}
 -\frac{1}{12}\frac{{\xi}\Omega_{str}}{u_{s}^{4}u_{t}^{2}}\}}
\end{eqnarray}
where $\beta$ is related to the bare QCD coupling constant,
$\xi=a_{s}/a_{t}$ is the aspect ratio for anisotropy (we take $\xi
=5$ in this work), $u_{s}$ and $u_{t}$ are the tadpole improvement
parameters of spatial and temporal gauge links, respectively.
$\Omega_{C}=\sum_{C}\frac{1}{3}Re Tr(1-W_{C})$, with $W_{C}$
denoting the path-ordered product of link variables along a closed
contour $C$ on the lattice. $\Omega_{sp}$ includes the sum over all
spatial plaquettes on the lattice, $\Omega_{tp}$ includes the
temporal plaquettes , $\Omega_{sr}$ denotes the product of link
variables about planar $2{\times}1$ spatial rectangular loops, and
$\Omega_{str}$ refers to the short temporal rectangles(one temporal
link, two spatial). Practically, $u_{t}$ is set to 1, and $u_{s}$ is
defined by the expectation value of the spatial plaquette,
$u_{s}=<\frac{1}{3}TrP_{ss^{'}}>^{1/4}$.

\subsection{Determination of critical temperature}
Since the temperature $T$ on the lattice is defined by
\begin{equation}
T=\frac{1}{N_t a_t},
\end{equation}
where $N_t$ is the temporal lattice size, $T$ can be changed by
varying either $N_t$ or the coupling constant $\beta$ which is
related directly to the lattice spacing. In order for the critical
temperature to be determined with enough precision, for a given
$N_t=24$, we first determine the critical coupling $\beta_c$,
because $\beta$ can be changed continuously. The order parameter is
chosen as the susceptibility $\chi_P$ of Polyakov line, which is
defined as
\begin{equation}
\chi_P =\langle\Theta ^{2}\rangle-\langle \Theta \rangle^{2}
\end{equation}
where $\Theta$ is the $Z(3)$ rotated Polyakov line,
\begin{eqnarray}
\Theta &=&\left\{ \
\begin{array}{ll}
{\rm Re}P\exp[-2\pi i/3];& \arg P\in \lbrack  \pi/3,\pi  ) \\
{\rm Re}P ;              & \arg P\in \lbrack -\pi/3,\pi/3) \\
{\rm Re}P\exp[ 2\pi i/3];& \arg P\in \lbrack -\pi, -\pi/3)
\end{array}
\right.,
\end{eqnarray}
and $P$ represents the trace of the spatially averaged Polyakov line
on each gauge configuration.

\begin{figure}
\centering
\includegraphics[height=5cm]{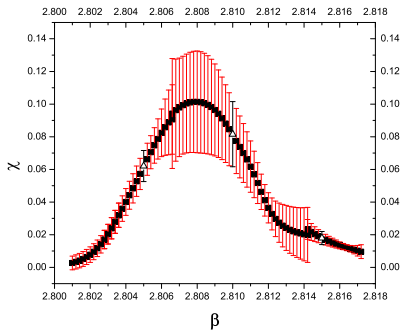}
\caption{\label{n_wt1} The $\chi_P$ extrapolation based on the
spectral density method. The open triangles denote the simulated
values of $\chi_P$, while the filled squares are the extrapolated
values. The peak position gives the critical $\beta_c=2.808$. }
\end{figure}

After a $\beta$-scanning on $L^4=24^4$ anisotropic lattices with
$\xi=5$, the critical point is trapped in a very narrow window
$\beta_c \in [2.800,2.820]$. In order to determine $T_c$ more
precisely, a more refined study is carried out in the $\beta$ window
mentioned above with much larger statistics through the spectral
density method. Practically, the spectral density
method~\cite{prl61, npb17} is applied to extrapolate the simulated
$\chi_P$'s at $\beta=2.805, 2.810$, and 2.815. In
table~\ref{beta:t1} are the numbers of heat-bath sweeps for each
$\beta$. The extrapolation results are illustrated in
Fig.~\ref{n_wt1} where the open triangles denote the simulated
values of $\chi_P$, while the filled squares are the extrapolated
values. Finally, the peak position gives the critical coupling
constant ${\beta}_{c}= 2.808$, which corresponds to the critical
temperature $T_c\approx 0.724 r_0^{-1}=296~{\rm MeV}$ with the
lattice spacing $r_0/a_s=3.476$~\cite{mpla21} and
$r_0^{-1}=410(20)\,{\rm MeV}$.

\begin{table}
\caption{\label{beta:t1} The simulation parameters for the
determination of the critical point. The configurations are selected
every ten sweeps.}
\begin{ruledtabular}
\begin{tabular}{cccccc}
$\beta$ & Total configurations & Thermalization & Bin size\\
\hline
2.80 & 20000 & 5000 & 1000 \\
2.805 & 30000 & 10000 & 1000 \\
2.81 & 30000 & 10000 & 1000 \\
2.815 & 20000 & 5000 & 1000 \\
2.82 & 8000 & 3000 & 500
\end{tabular}
\end{ruledtabular}
\end{table}

With $T_c$ fairly determined, the working coupling constant $\beta$
is set based on two requirements. First, the spatial volume of the
lattice should be large enough in order for the glueballs to be free
of any sizable finite volume effects. Secondly, we require that
temporal lattice has a good resolution even at the temperature
$T\sim 2T_c$. Practically the working coupling constant is finally
set to be $\beta=3.2$. The lattice spacing at this $\beta$ is set by
calculating the static potential $V(r)$ on an anisotropic lattice
$24^3\times 128$. With the conventional parametrization of V(r),
\begin{equation}
V(r)=V_0+\sigma r +\frac{e_c}{r},
\end{equation}
the lattice spacing $a_s$ is determined in the units of $r_0$ to be
\begin{equation}
\frac{a_s}{r_0} = \sqrt{\frac{\sigma a_s^2}{1.65+e_c}}=0.1825(7)
\end{equation}
where $r_0$ is the hadronic scale parameter. If we take
$r_0^{-1}=410(20)\,{MeV}$, we have $a_s=0.0878(4)\,{\rm fm}$. The
spatial volume at $L=24$ is therefore estimated to be $(2.1\,{\rm
fm})^3$. On the other hand, using $T_c=296$MeV obtained at
$\beta=2.808$ as a rough estimate of $T_c$ and ignoring the
systematic error due to finite lattice spacings, $T_c$ and $2T_c$ at
$\beta=3.2$ are expected to be achieved around $N_t \sim 40 $ and
$N_t \sim 20$, respectively. Obviously, the above two requirements
are all satisfied.
\begin{table}
\caption{\label{pl's error}Listed are the parameters used to check
the critical behavior for $\beta$=3.2. The configurations are
selected every ten sweeps. }
\begin{ruledtabular}
\begin{tabular}{ccccccc}
  $N_t$ & Total configurations & Thermalization & $<P>$ & $\chi_P$   \\
  \hline
   60    &   2000    &   500     &   -8.73$\times 10^{-5}$  &   6.65$\times 10^{-5}$    \\
   48    &   2000    &   500     &   6.01$\times 10^{-5}$   &   1.81$\times 10^{-4}$    \\
   44    &   8000    &   2000    &   2.25$\times 10^{-3}$   &   3.12$\times 10^{-3}$    \\
   40    &   8000    &   2000    &   1.72$\times 10^{-2}$   &   9.14$\times 10^{-3}$    \\
   36    &   8000    &   2000    &   5.21$\times 10^{-2}$   &   3.10$\times 10^{-3}$    \\
   32    &   3000    &   1000    &   8.51$\times 10^{-2}$   &   2.23$\times 10^{-3}$    \\
   28    &   2000    &   500     &   0.1253      &   2.00$\times 10^{-3}$    \\
   24    &   2000    &   500     &   0.1817      &   2.09$\times 10^{-3}$    \\
   20    &   2000    &   500     &   0.2571      &   1.82$\times 10^{-3}$
\end{tabular}
\end{ruledtabular}
\end{table}

\begin{figure}
\centering
\includegraphics[height=5cm]{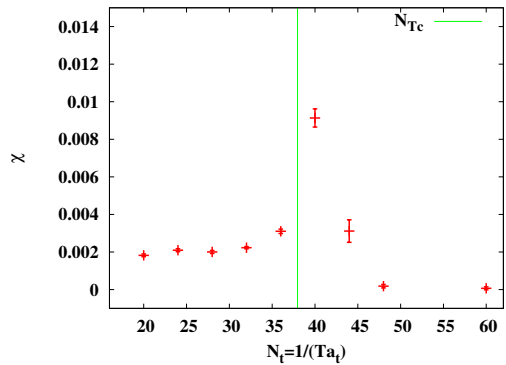}
\caption{$\chi_P$ is plotted versus $N_t$ at $\beta=3.2$. There is a
peak of $\chi_P$ near $N_t=40$.  \label{fig_chi}}
\end{figure}

Based on the discussions above, with a fixed $\beta=3.2$, the
calculations of the thermal correlators of glueballs are carried out
on a series of lattice $24^3\times N_t$ with $N_{t}=$ 20, 24, 28,
32, 36, 40, 44, 48, 60, 80, and 128, which cover the temperature
range $0.3T_c<T<2T_c$. As a cross-check, $\chi_P$ at different $N_t$
are calculated first and the results are shown in Fig.~\ref{fig_chi}
and Table~\ref{pl's error}. It is clear that the expectation value
of the Polyakov line drops to zero near $N_t = 40$ and the peak
position of $\chi_P$, which gives the critical temperature, is
trapped between $N_t=36$ and $N_t=40$. In practice, we do not carry
out a precise determination of $T_c$ at $\beta=3.2$, but take the
temperature at $N_t=38$,
$T\approx(38a_t)^{-1}=(38a_s/\xi)^{-1}=296$MeV, as an approximation
of $T_c~(\beta=3.2)$, to scale the temperatures involved in this
work. It should be noted that, owing to the lattice artifact, the
critical temperature $T_c$ determined at different lattice spacing
(or $\beta$) may differ from each other. The closeness of
$T_c(\beta=2.808)$ and $T_c(\beta=3.2)$ may signal that the lattice
spacing dependence of $T_c$ is mild in this work due to the
application of the improved gauge action.

\subsection{Variational method}
It is known that many states contribute to a hadronic two-point
function. Ideally one can extract the information of the
lowest-lying states from the two-point function in the large time
region if it lasts long enough in the time direction. This is the
case for some light hadron states, such as $\pi$ meson, $K$ meson,
etc. However, for heavy particles, especially for glueballs whose
correlation function are much more noisy than that of conventional
hadrons made up of quarks, their two-point functions damp so fast
with time that they are always undermined by noise rapidly before
the ground states dominate. Practically, in the study of the
glueball sector, in order to enhance the overlap of the glueball
operators to the ground state, the commonly used techniques are the
smearing schemes and the variational techniques. In this work, we
adopt the sophisticated strategy implemented by the studies of the
zero-temperature glueball spectrum~\cite{prd56,prd60,prd73}, which
is outlined below.
\begin{figure}
\includegraphics[height=5.0cm]{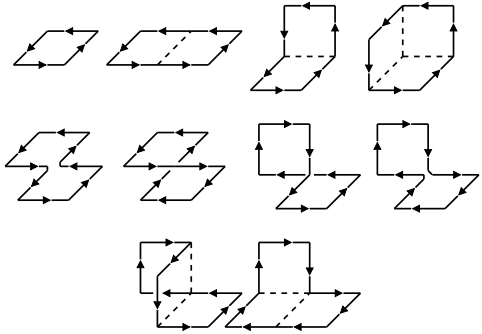}
\caption{\label{proto} Prototype Wilson loops used in making the
smeared glueball operators\cite{prd60}.}
\end{figure}
First, for each gauge configuration, we perform six smearing/fuzzing
schemes to the spatial links, which are various combinations of the
single-link procedure (smearing) and the double-link procedure
(fuzzing)
\begin{widetext}
\begin{eqnarray}  \label{s/f}
U_j^s(x) &=& P_{SU(3)}\{U_j(x)+\lambda_s\sum\limits_{\pm(k\neq j)}
U_k(x)U_j(x+\hat{k})U_k^{\dagger}(x+\hat{j})\},\nonumber \\
U_j^f(x) &=&
P_{SU(3)}\{U_j(x)U_j(x+\hat{j})+\lambda_f\sum\limits_{\pm(k\neq
j)}U_k(x)U_j(x+\hat{k})U_j(x+\hat{j}+\hat{k})U_k(x+2\hat{j})\},
\end{eqnarray}
\end{widetext}
where $P_{SU(3)}$ denotes the projection into $SU(3)$ and is
realized by the Jacobi method~\cite{liu2}. The six schemes are given
explicitly as $ s_{\lambda_s}^{10}$, $ s_{\lambda_s}^{18}$, $
s_{\lambda_s}^{26}$, $ f_{\lambda_f}\bigotimes s_{\lambda_s}^{10}$,
$ f_{\lambda_f}\bigotimes s_{\lambda_s}^{18}$, $
f_{\lambda_f}\bigotimes s_{\lambda_s}^{26}$, where $s/f$ denotes the
smearing/fuzzing procedure defined in Eq.~(\ref{s/f}), and
$\lambda_s/\lambda_f$ the tunable parameter which we take $\lambda_s
= 0.1$ and $ \lambda_f = 0.5$ in this work. Secondly, we choose the
same prototype Wilson loops as that in Ref.~\cite{prd60} (as shown
in Fig.~\ref{proto}), such that for each smearing/fuzzing scheme,
all the different spatially oriented copies of these prototypes are
calculated from the smeared gauge configurations. Thus for a given
irreducible representation $R$ of the spatial symmetry group $O$,
say, $R=A_1, A_2, E, T_1$, or $T_2$, a realization of $R$ can be a
specific combination of differently oriented Wilson loops generated
from the same prototype loop (one can refer to Ref.~\cite{prd73} for
the concrete combinational coefficients). The glueball operators
$\phi$ with the quantum number $R^{PC}$ are thereby constructed
along with the spatial reflection and the time inversion operations.
In practice, we establish four realizations of each $R^{PC}$ which
are based on four different prototypes, respectively. Therefore,
along with the six smearing/fuzzing schemes, an operator set of the
same specific quantum number $R^{PC}$ is composed of 24 different
operators, $\{\phi_\alpha, \alpha = 1,2,\ldots, 24\}$. The last step
is the implementation of the variational method (VM). The main goal
of VM is to find an optimal combination of the set of operators,
$\Phi=\sum v_{\alpha}\phi_\alpha$, which overlaps most to a specific
state (in this work, we only focus on the ground states). The
combinational coefficients ${\bf v}=\{v_\alpha, \alpha=1,2,\ldots,
n\}$ can be obtained by minimizing the effective mass,
\begin{equation}
\label{mass} \tilde{m}(t_D)=-\frac{1}{t_D}\ln\frac
{\sum\limits_{\alpha\beta} v_\alpha
v_\beta\tilde{C}_{\alpha\beta}(t_D)} {\sum\limits_{\alpha\beta}
v_\alpha v_\beta \tilde{C}_{\alpha\beta}(0)},
\end{equation}
at $t_D=1$, where $\tilde{C}_{\alpha\beta}(t)$ is the correlation
matrix of the operator set,
\begin{equation}
\tilde{C}_{\alpha\beta}(t) = \sum\limits_{\tau}\langle
0|{\phi}_\alpha(t+\tau){\phi}_\beta(\tau)|0\rangle.
\end{equation}
This is equivalent to solving the generalized eigenvalue equation
\begin{equation}
\label{eigen} \tilde{C}(t_D){\bf v}^{(R)} =
e^{-t_D\tilde{m}(t_D)}\tilde{C}(0){\bf v}^{(R)},
\end{equation}
and the eigenvector ${\bf v}$ gives the desired combinational
coefficients. Thus, the optimal operator that couples most to a
specific states (the ground state in this work) can be built up as
\begin{equation}
\Phi = \sum\limits_\alpha v_\alpha \phi_\alpha,
\end{equation}
whose correlator $C(t)$ is expected to be dominated by the
contribution of this state.

\section{DATA ANALYSIS OF THE THERMAL CORRELATORS OF
GLUEBALLS}

All 20 $R^{PC}$ channels, with $R=A_1, A_2, E,T_1, T_2$ and
$PC=++,+-,-+,--$, are considered in the calculation of the thermal
correlators of glueballs on anisotropic lattices mentioned in Sec.
II. At each temperature, after 10000 pseudo-heat-bath sweeps of
thermalization, the measurements are carried out every three
compound sweeps, with each compound sweep composed of one
pseudo-heat-bath and five micro-canonical over-relaxation(OR)
sweeps. In order to reduce the possible autocorrelations, the
measured data are divided into bins of the size $n_{\rm mb}=400$,
and each bin is regarded as an independent measurement in the data
analysis procedure. The numbers of bins $N_{\rm bin}$ and $n_{mb}$
at various temperature are listed in Table~\ref{simu_para}.
\begin{table}
\caption{\label{simu_para}Simulation parameters to calculate
glueball spectrum. $\beta=3.2$, $a_s=0.0878\,{\rm fm}$,
$L_s=2.11\,{\rm fm}$.}
\begin{ruledtabular}
\begin{tabular}{cccc}
   $N_t$    &   $T/T_{c}$   &   $n_{\rm mb}$    &   $N_{\rm bin}$   \\
  \hline
    128     &   0.30        &   400          &   24         \\
    80      &   0.47        &   400          &   30         \\
    60      &   0.63        &   400          &   44         \\
    48      &   0.79        &   400          &   40         \\
    44      &   0.86        &   400          &   44         \\
    40      &   0.95        &   400          &   40         \\
   \hline
    36      &   1.05        &   400          &   40         \\
    32      &   1.19        &   400          &   56         \\
    28      &   1.36        &   400          &   40         \\
    24      &   1.58        &   400          &   40         \\
    20      &   1.90        &   400          &   40
\end{tabular}
\end{ruledtabular}
\end{table}

Theoretically, under the periodic boundary condition in the temporal
direction, the temporal correlators $C(t,T)$ at the temperature $T$
can be written in the spectral representation as
\begin{eqnarray}
\label{aaa} C(t,T)&\equiv& \frac{1}{Z(T)}{\rm Tr} \left( e^{-
H/T}\Phi(t)\Phi(0)\right)\nonumber \\
&=&\sum\limits_{m,n}\frac{|\langle n|\Phi|m\rangle|^2}
{2Z(T)} \exp \left(-\frac{E_m+E_n}{2T}\right)\nonumber \\
&&\times
\cosh\left[\left(t-\frac{1}{2T}\right)(E_n-E_m)\right]\nonumber\\
&=&\int\limits_{-\infty}^{\infty} d\omega \rho(\omega) K(\omega,T),
\end{eqnarray}
with a $T$-dependent kernel
\begin{equation}
K(\omega,T)=\frac{\cosh(\omega/(2T)-\omega t)}{\sinh(\omega/(2T))}
\end{equation}
and the spectral function,
\begin{eqnarray}
\label{spectral}
 \rho(\omega)&=& \sum\limits_{m,n}\frac{|\langle
n|\Phi|m\rangle|^2}{2Z(T)}e^{-E_m/T}\nonumber\\
&\times& (\delta(\omega-(E_n-E_m)-\delta(\omega-(E_m-E_n)),
\end{eqnarray}
where $Z(T)$ is the partition function at $T$, and $E_n$ the energy
of the thermal state $|n\rangle$ ($|0\rangle$ represents the vacuum
state). In the zero-temperature limit($T\rightarrow 0$), due to the
factor $\exp(-E_m/T)$, the spectral function $\rho(\omega)$
degenerates to
\begin{equation}
\label{single_pole} \rho(\omega)=\sum\limits_n\frac{|\langle
0|\Phi|n\rangle|^2}{2Z(0)}
\left(\delta(\omega-E_n)-\delta(\omega+E_n)\right),
\end{equation}
thus we have the function form of the correlation function,
\begin{equation}
C(t,T=0)=\sum\limits_n W_n e^{-E_n \tau}
\end{equation}
with $W_n=|\langle 0|\Phi|n\rangle|^2/Z(0)$.

However, for any finite temperature (this is always the case for
finite lattices), all the thermal states with the nonzero matrix
elements $\langle m|\Phi|n\rangle$ may contribute to the spectral
function $\rho(\omega)$. Intuitively in the confinement phase, the
fundamental degrees of freedom are hadronlike modes, thus the
thermal states should be multihadron states. If they interact weakly
with each other, we can treat them as free particles at the lowest
order approximation and consider $E_m$ as the sum of the energies of
hadrons including in the thermal state $|m\rangle$. Since the
contribution of a thermal state $|m\rangle$ to the spectral function
is weighted by the factor $\exp(-E_m/T)$, apart from the vacuum
state, the maximal value of this factor is $\exp(-M_{\rm min}/T)$
with $M_{\rm min}$ the mass of the lightest hadron mode in the
system. As far as the quenched glueball system is concerned, the
lightest glueball is the scalar, whose mass at the low temperature
is roughly $M_{0^{++}}\sim 1.6$ GeV, which gives a very tiny weight
factor $\exp(-M_{0^{++}}/T_c)\sim 0.003$ at $T_c$ in comparison with
unity factor of the vacuum state. That is to say, for the quenched
glueballs, up to the critical temperature $T_c$, the contribution of
higher spectral components beyond the vacuum to the spectral
function are much smaller than the statistical errors (the relative
statistical errors of the thermal glueball correlators are always a
few percent) and can be neglected. As a result, the function form of
$\rho(\omega)$ in Eq.~\ref{single_pole} can be a good approximation
for the spectral function of glueballs at least up to $T_c$.
Accordingly, considering the finite extension of the lattice in the
temporal direction, the function form of the thermal correlators can
be approximated as
\begin{equation}
\label{cosh_function} C(t,T)=\sum\limits_n W_n
\frac{\cosh(M_n(1/(2T)-t))}{\sinh(M_n/(2T))},
\end{equation}
which is surely the commonly used function form for the study of
hadron masses at low temperatures on the lattice. As is always done,
the glueball masses $M_n$ derived by this function are called the
pole masses in this work.

\subsection{Results of the single-cosh fit}

Even though the above discussion are based on the weak-interaction
approximation for the hadronlike modes below $T_c$, we would like to
apply Eq.~\ref{cosh_function} to analyzing the thermal correlators
all over the temperature in concern. The interest of doing so is
twofold. First, the thermal scattering of the glueball-like modes
would result in a mass shift, say the deviation of the pole mass
from the glueball mass at zero-temperature, which reflects the
strength of the interaction at different temperature. Secondly, the
breakdown of this function form would signal the dominance of new
degrees of freedom instead of the hadronlike modes in the thermal
states.

In practice, after the thermal correlators $C(t, T)$ of the optimal
operators are obtained according to the steps described in Sec.
II(B), the pole masses of the ground state (or the lowest spectral
component) can be extracted straightforwardly. First, for each
$R^{PC}$ channel and at each temperature $T$, the effective mass
$M_{\rm eff}(t)$ as a function of $t$ is derived by solving the
equation
\begin{equation}
\frac{C(t+1,T)}{C(t,T)}=\frac{\cosh ( (t + 1 - N_{t}/2)a_{t}M_{\rm
eff}(t) )}{\cosh ( (t - N_{t}/2)a_{t}M_{\rm eff}(t) )}, \label{cosh}
\end{equation}
Secondly, the effective masses are plotted versus $t$ and the
plateaus give the fit windows $[t_1,t_2]$. Finally, the pole masses
of the ground states are obtained by fitting $C(t,T)$ through a
single-cosh function form. As a convention in this work, we use
$M_{G}$ to represent the mass of a glueball state in the physical
units and $M$ to represent the dimensionless mass parameter in the
data processing with the relation $M = M_{G}a_{t}$.
\begin{figure}[thb]
  \begin{center}
    \begin{tabular}{cc}
      \resizebox{40mm}{!}{\includegraphics{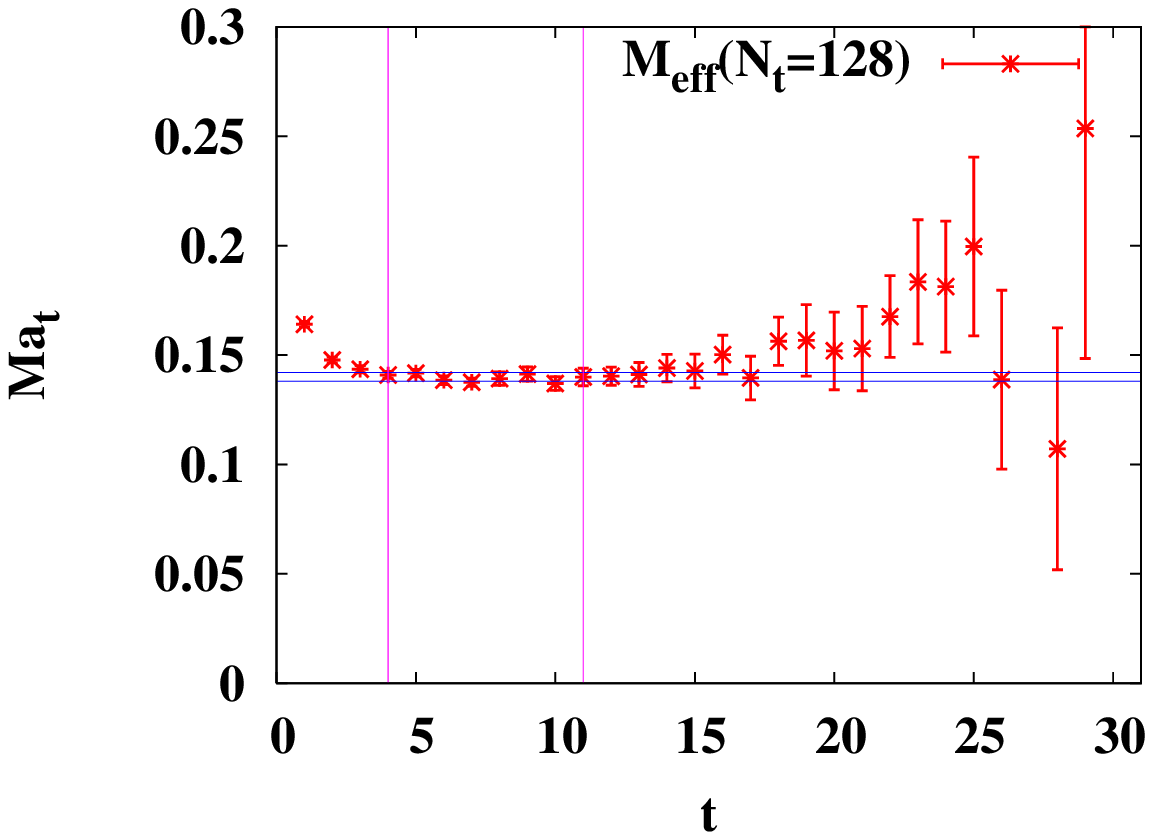}} &
      \resizebox{40mm}{!}{\includegraphics{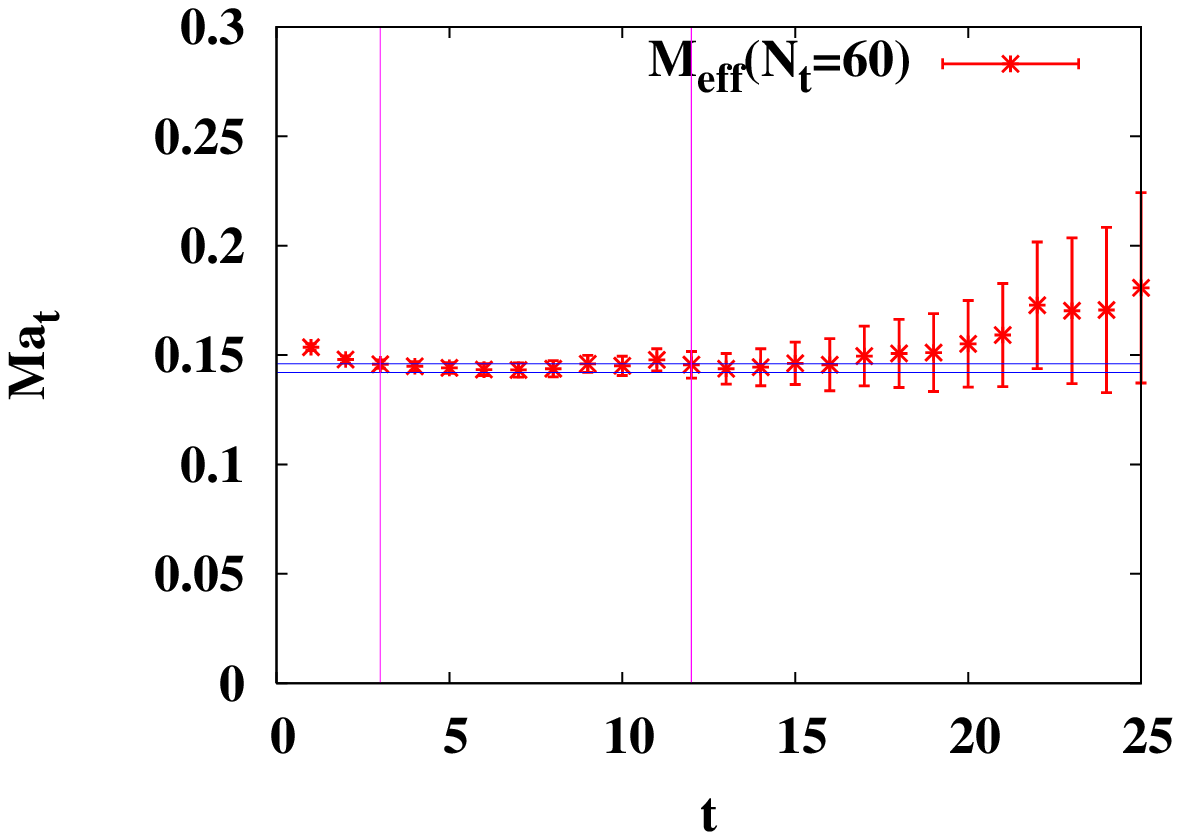}} \\
      \resizebox{40mm}{!}{\includegraphics{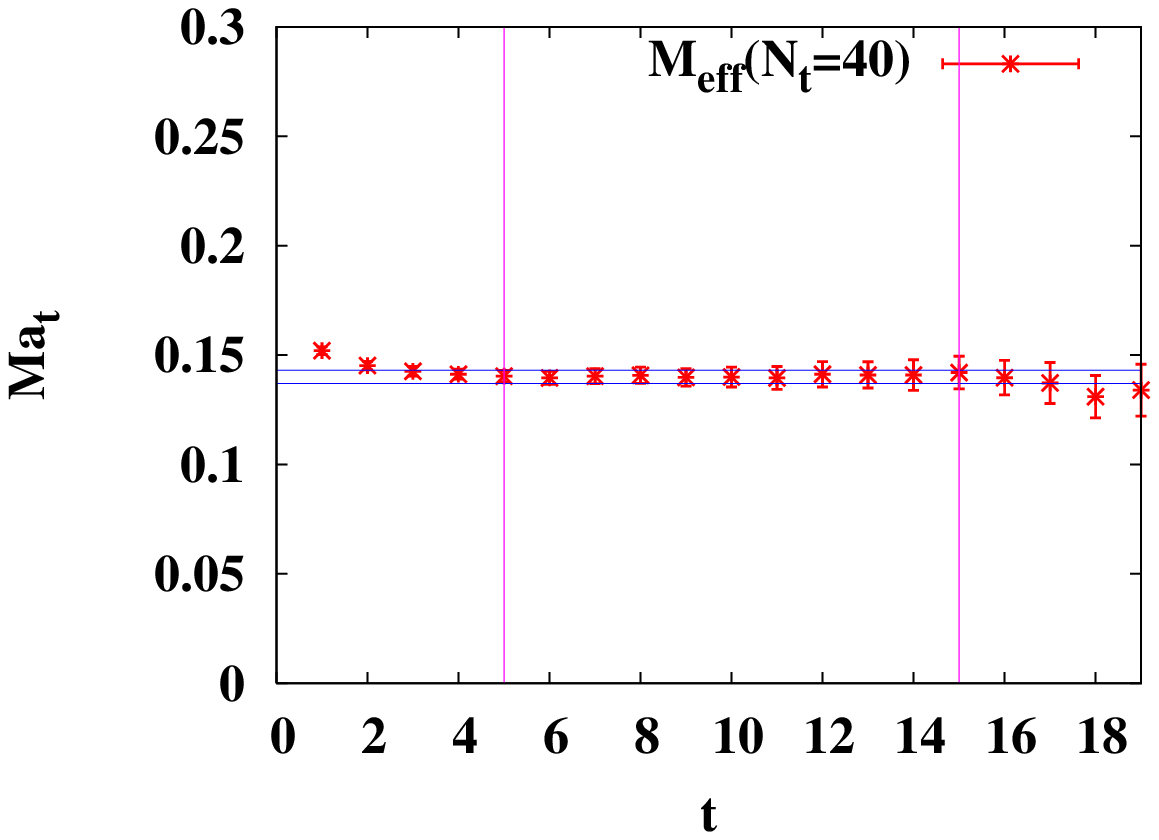}} &
      \resizebox{40mm}{!}{\includegraphics{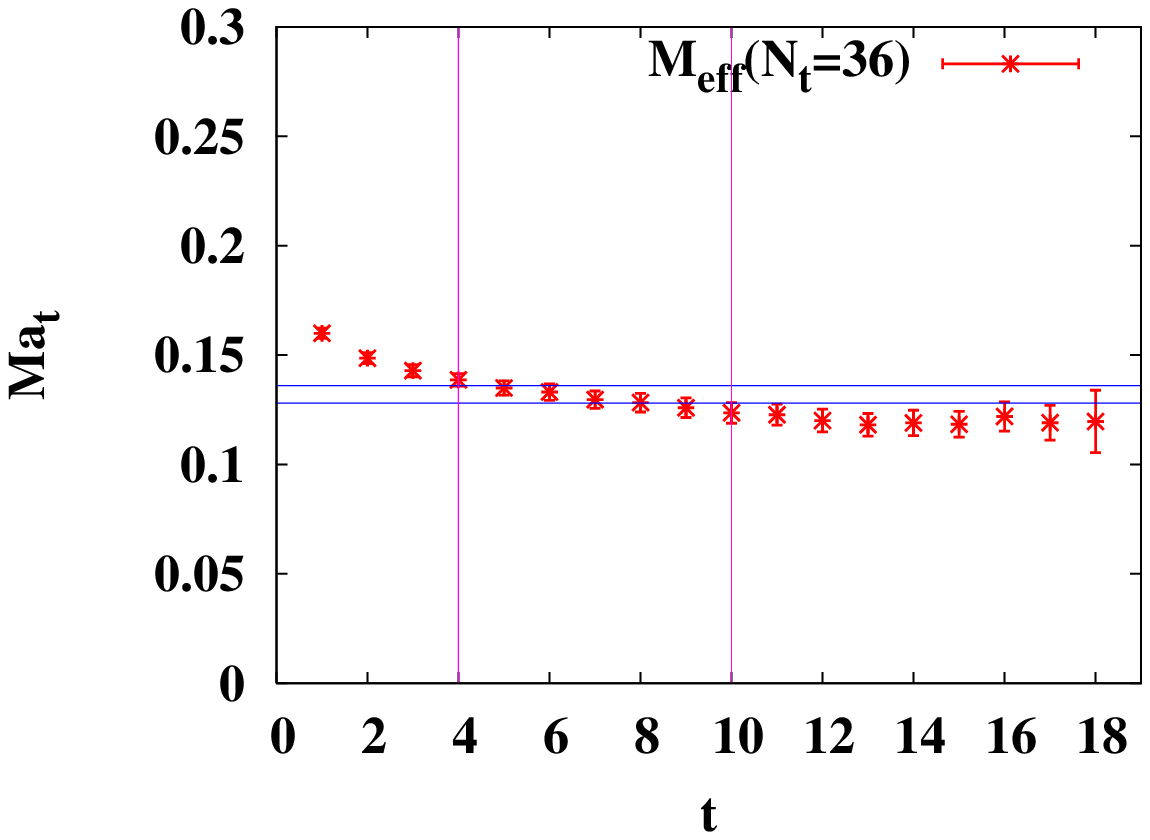}} \\
      \resizebox{40mm}{!}{\includegraphics{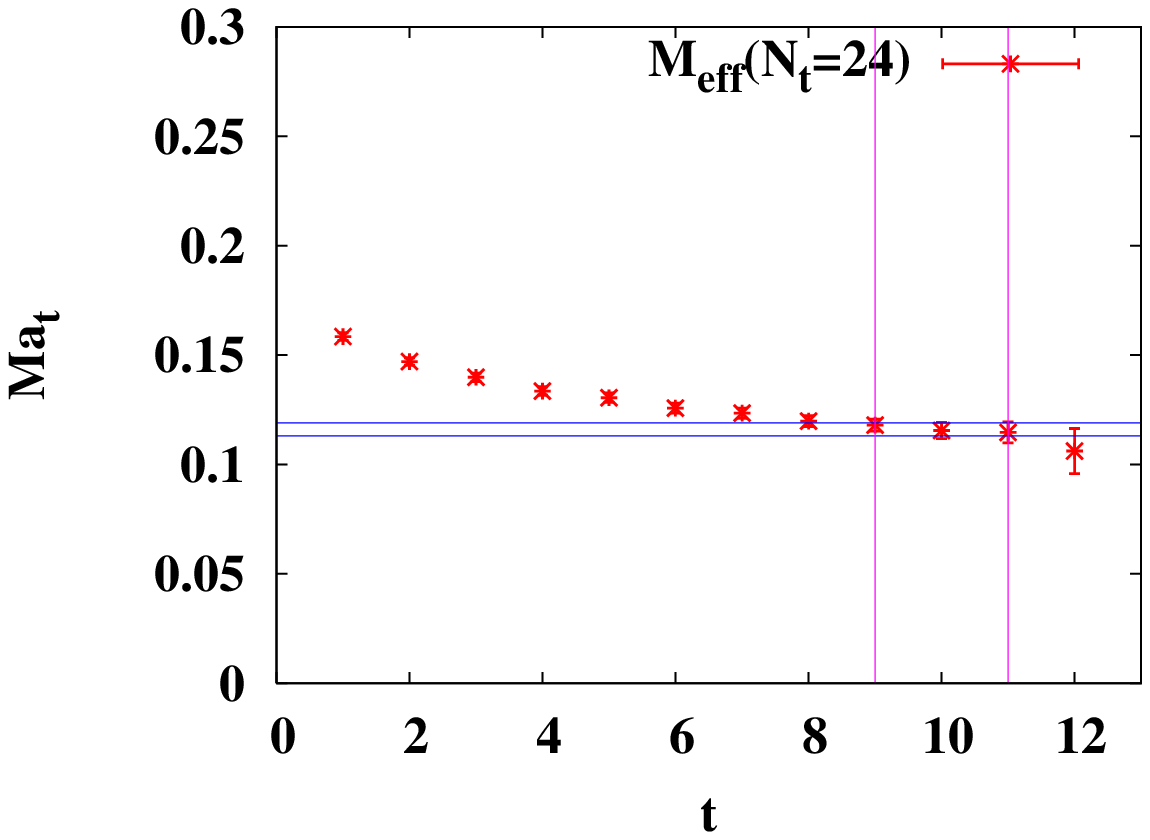}} &
      \resizebox{40mm}{!}{\includegraphics{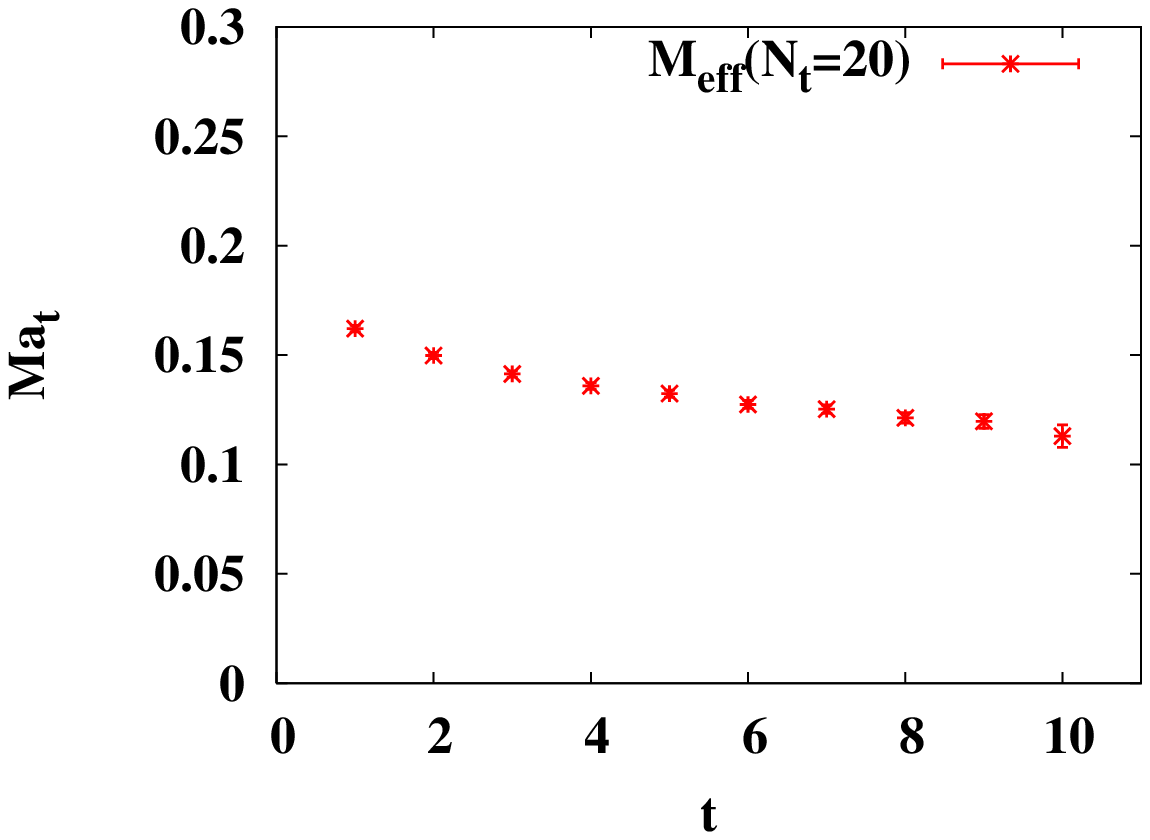}} \\
    \end{tabular}
    \caption{Effective masses at different temperatures in $A_1^{++}$ channel.
    Data points are the effective masses with jackknife error bars.
    The vertical lines indicate the time window $[t_1, t_2]$ over which the
    single-cosh fittings are carried out, while the horizontal lines illustrate
    the best-fit result of pole masses (in each panel the double horizontal lines
    represent the error band estimated by jackknife analysis)}
    \label{opt01}
  \end{center}
\end{figure}

\begin{figure}[thb]
  \begin{center}
    \begin{tabular}{cc}
      \resizebox{40mm}{!}{\includegraphics{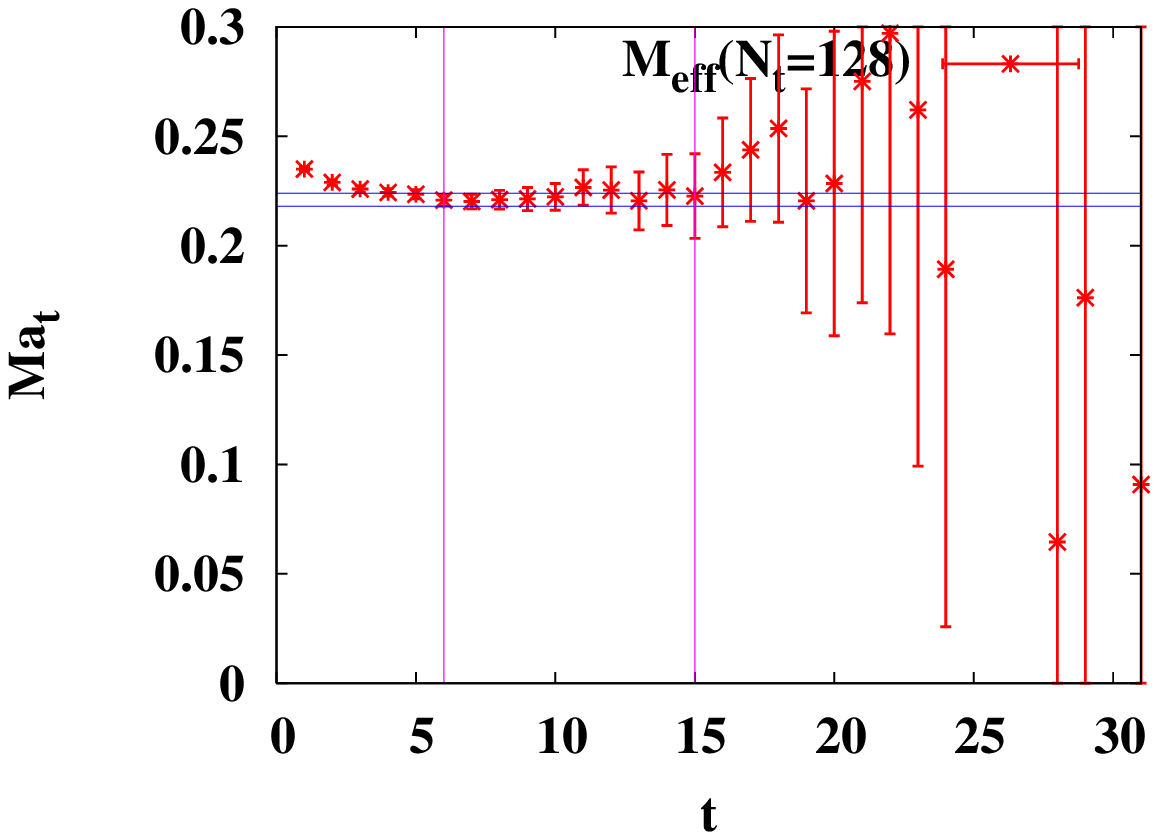}} &
      \resizebox{40mm}{!}{\includegraphics{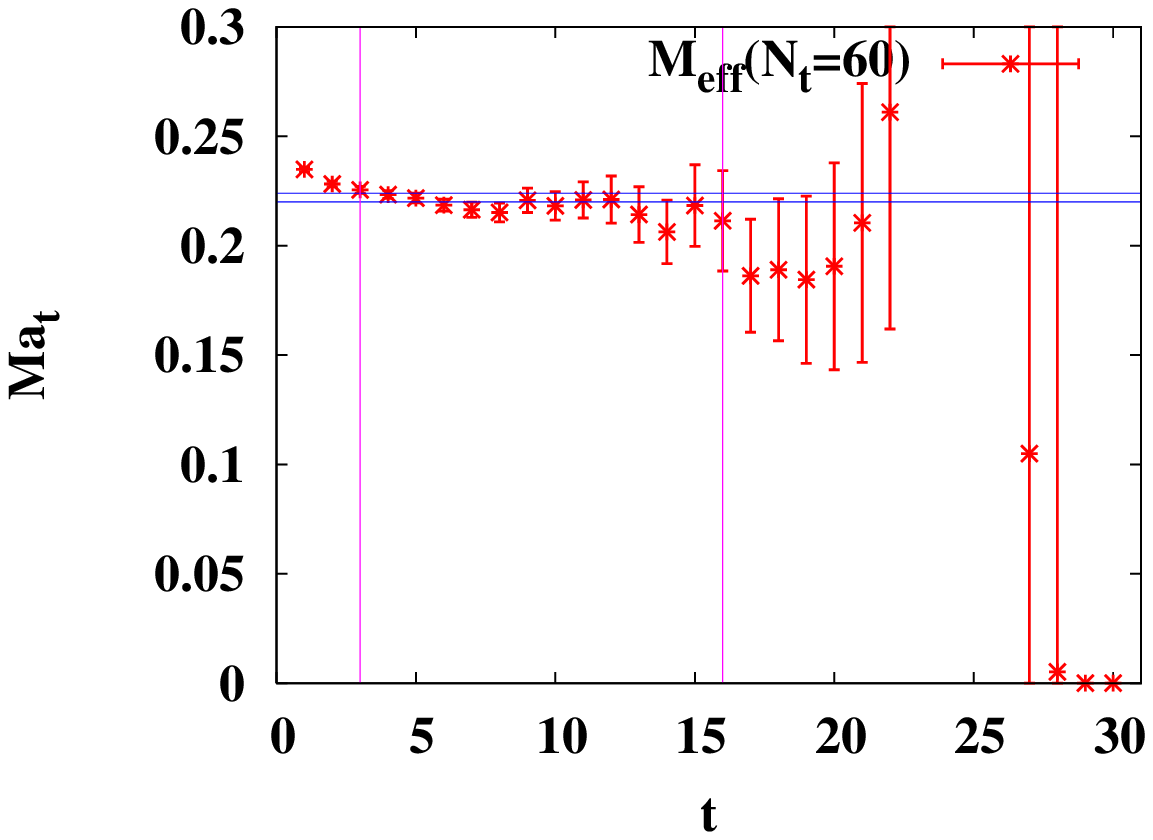}} \\
      \resizebox{40mm}{!}{\includegraphics{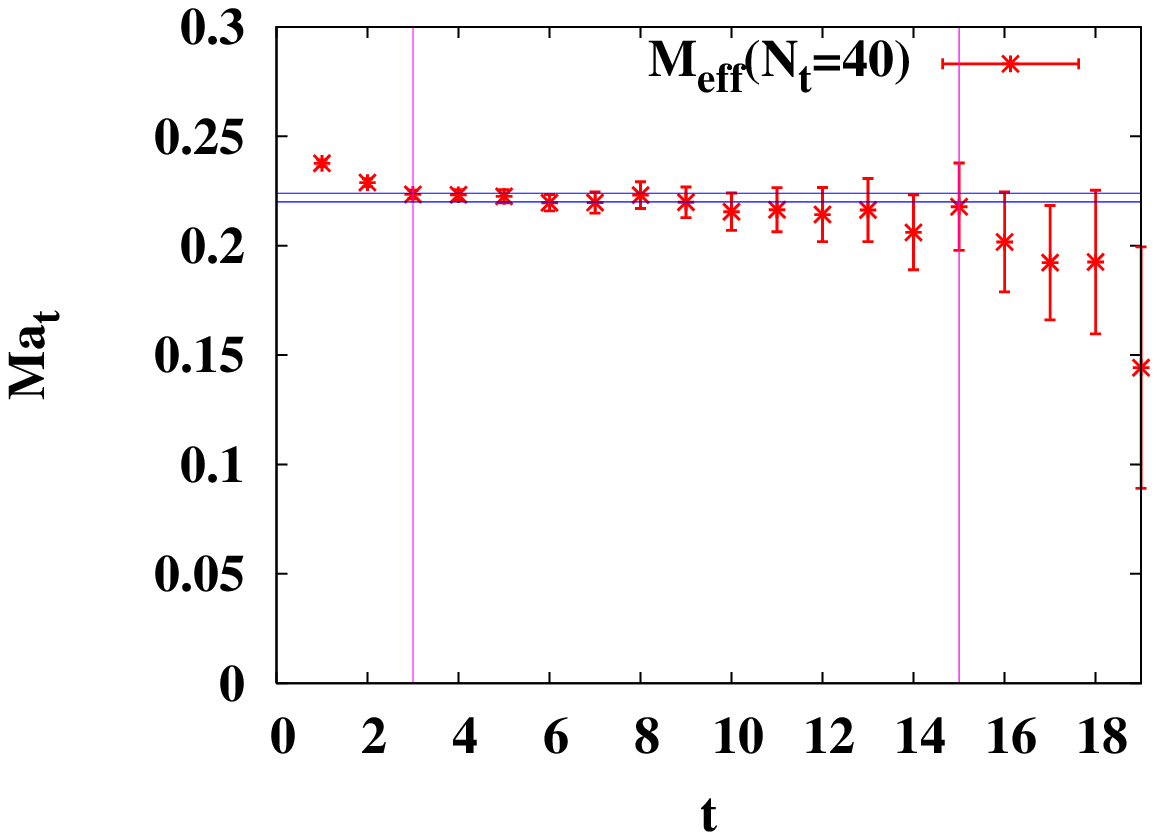}} &
      \resizebox{40mm}{!}{\includegraphics{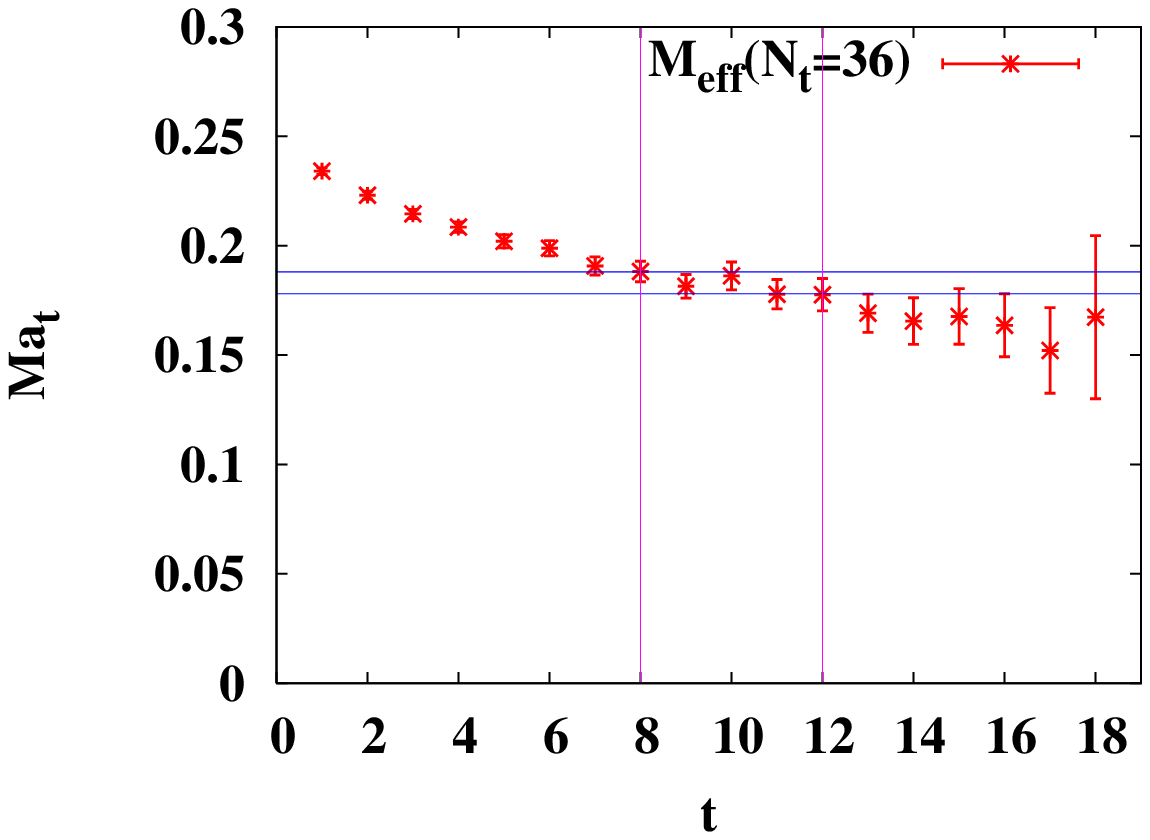}} \\
      \resizebox{40mm}{!}{\includegraphics{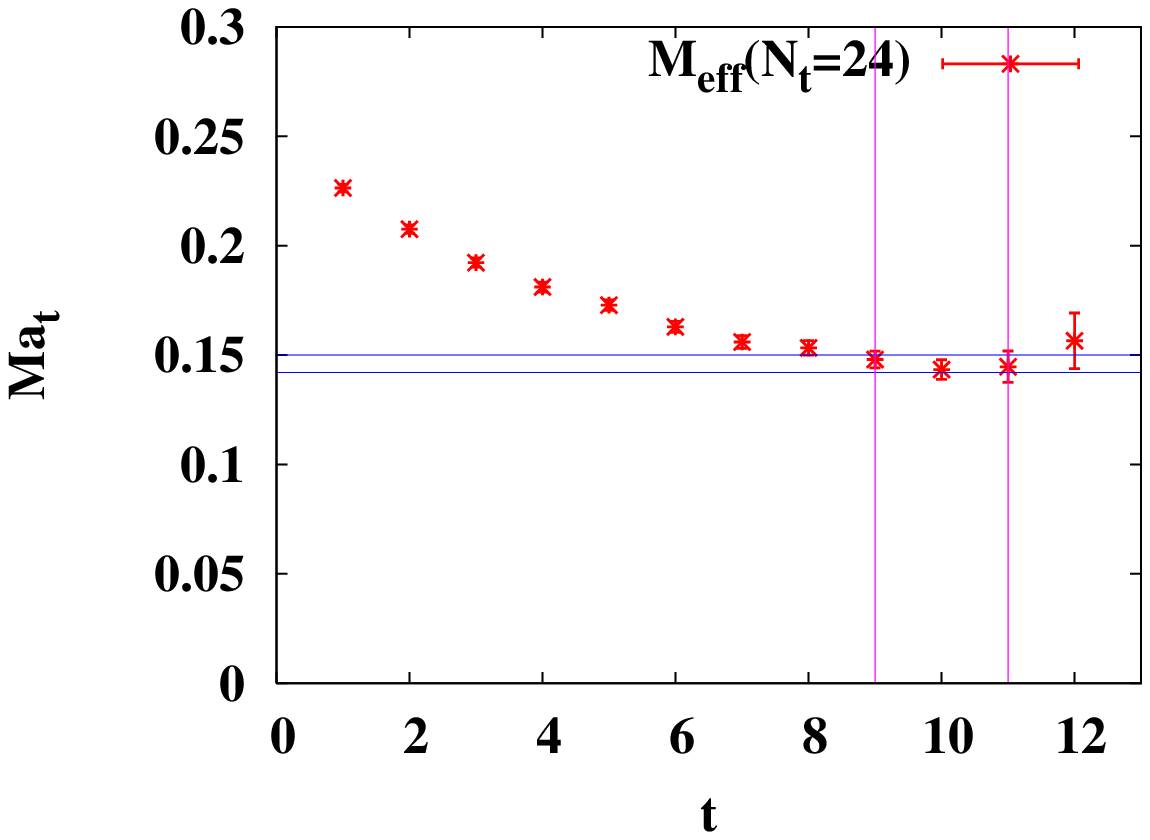}} &
      \resizebox{40mm}{!}{\includegraphics{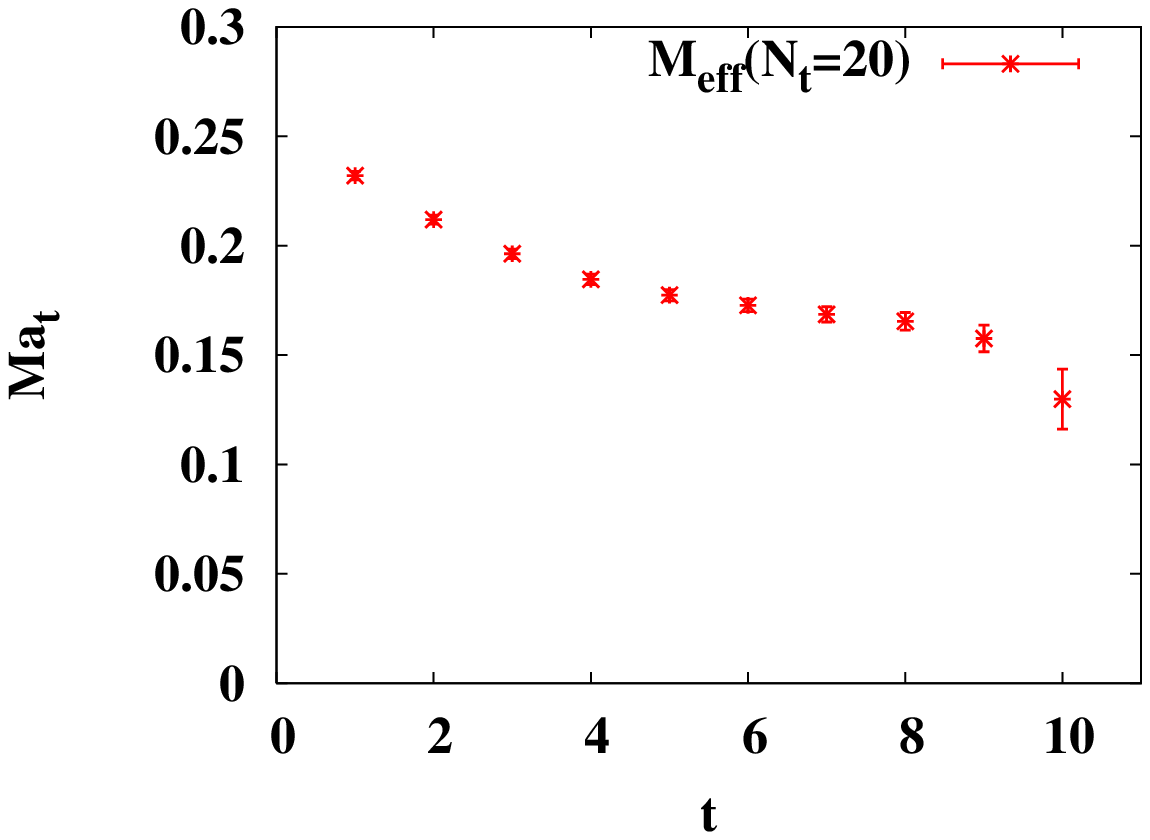}} \\
    \end{tabular}
    \caption{Similar to Fig.~\ref{opt01}, but in $A_{1}^{-+}$ channel.}
    \label{opt03}
  \end{center}
\end{figure}

\begin{figure}[thb]
  \begin{center}
    \begin{tabular}{cc}
      \resizebox{40mm}{!}{\includegraphics{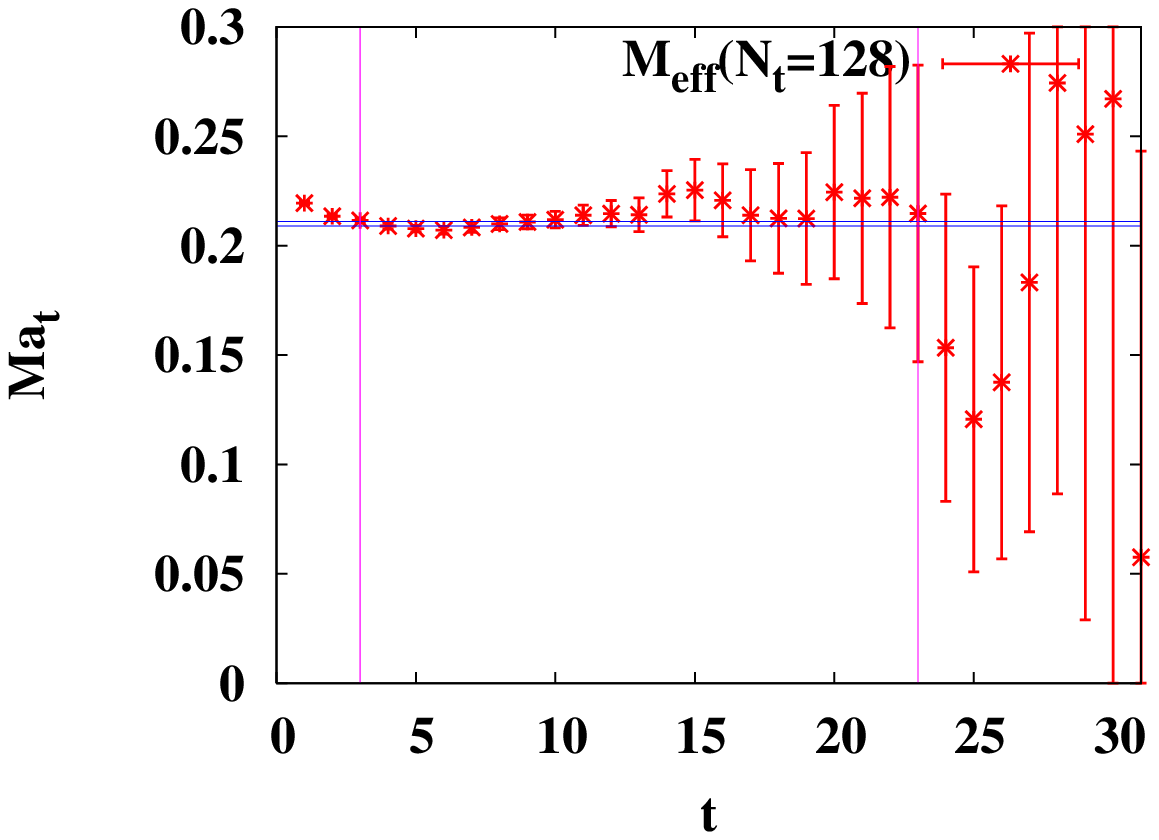}} &
      \resizebox{40mm}{!}{\includegraphics{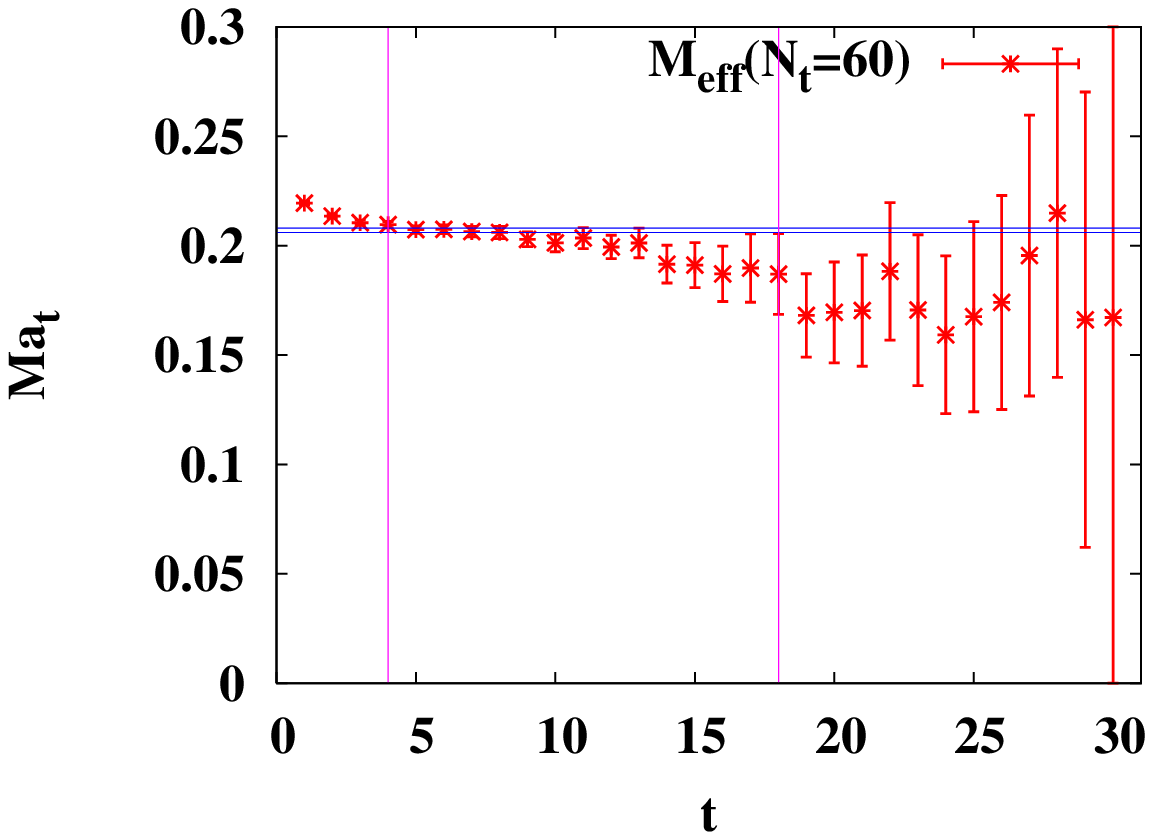}} \\
      \resizebox{40mm}{!}{\includegraphics{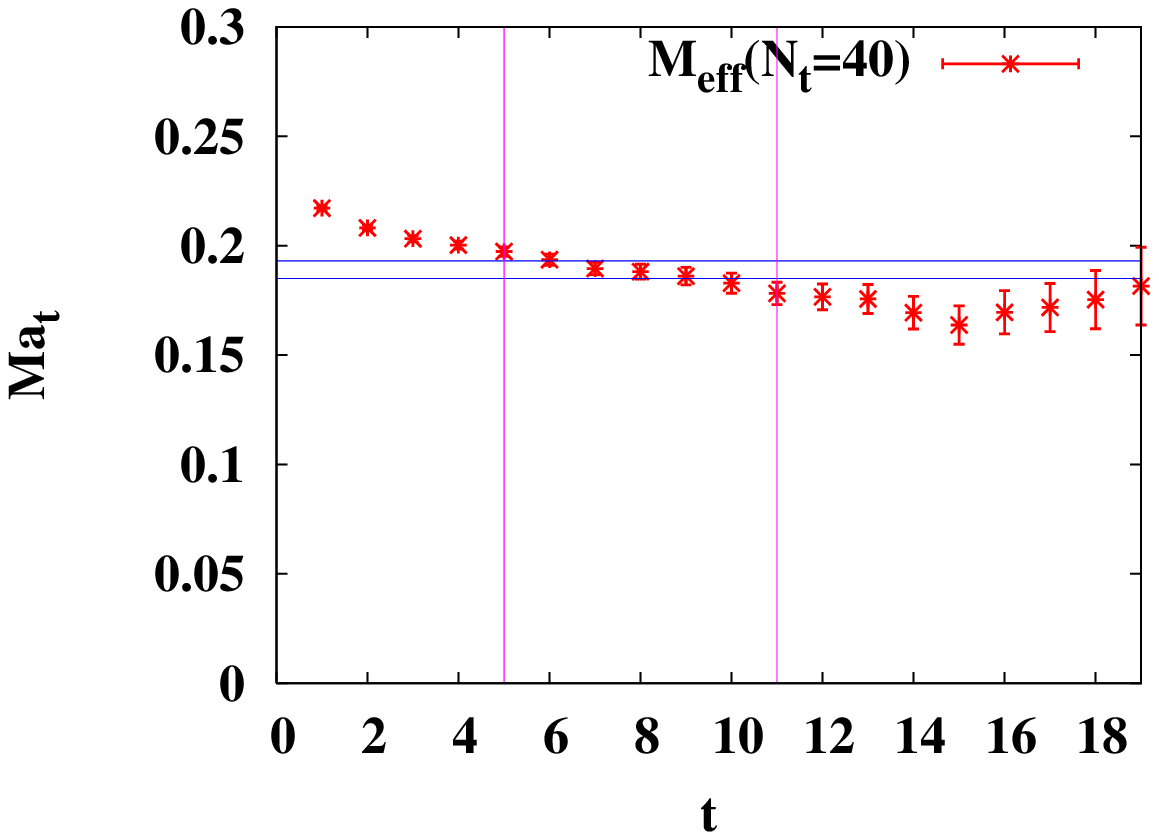}} &
      \resizebox{40mm}{!}{\includegraphics{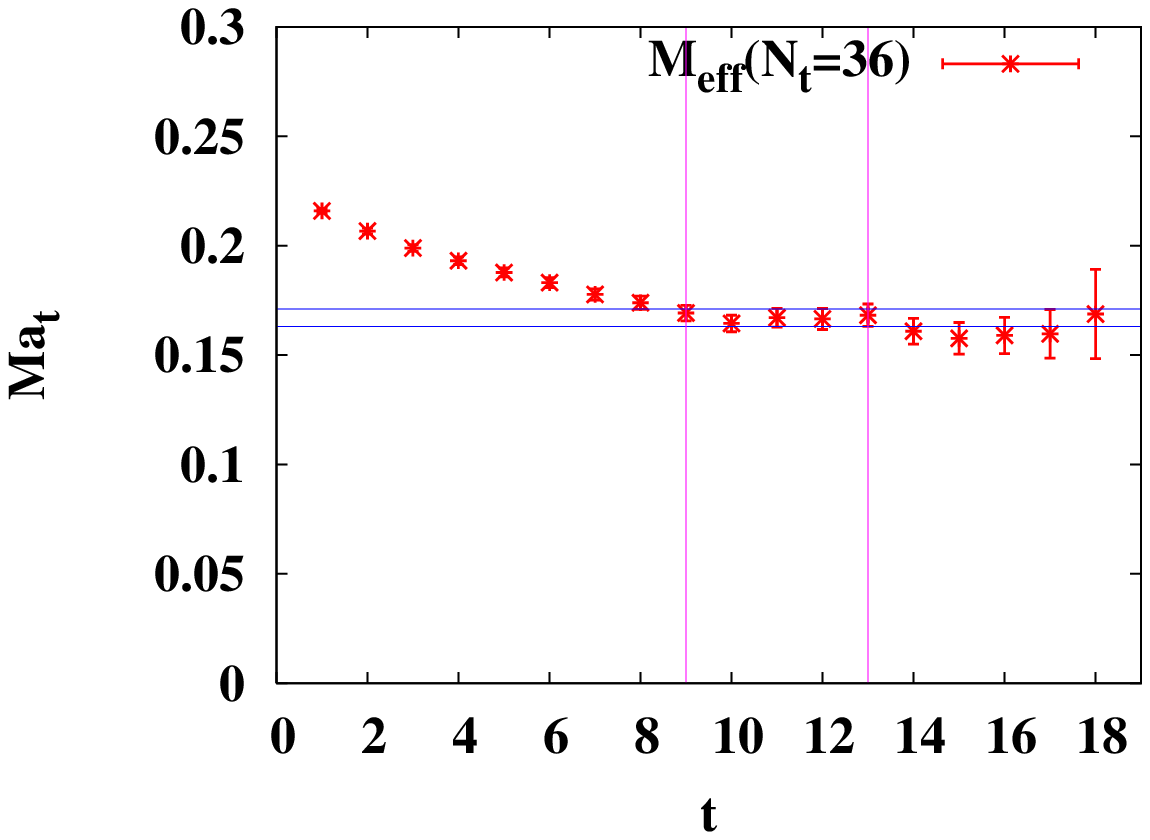}} \\
      \resizebox{40mm}{!}{\includegraphics{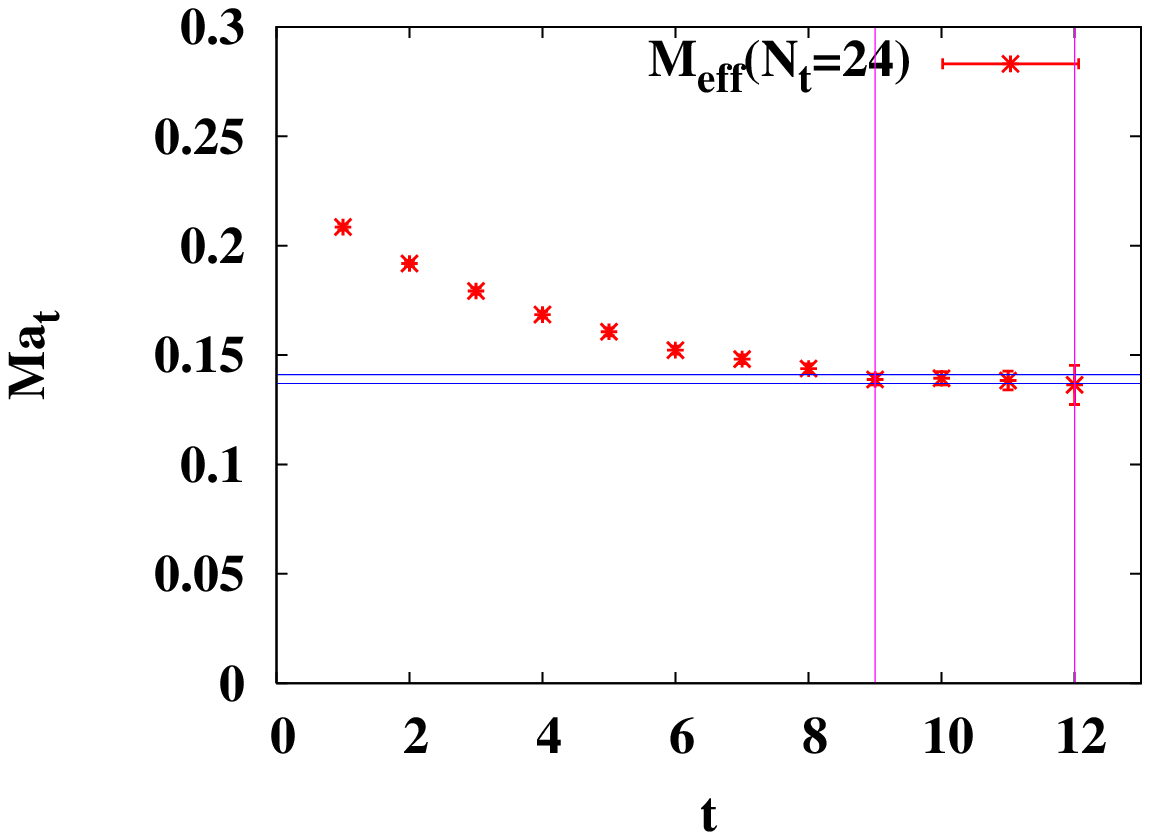}} &
      \resizebox{40mm}{!}{\includegraphics{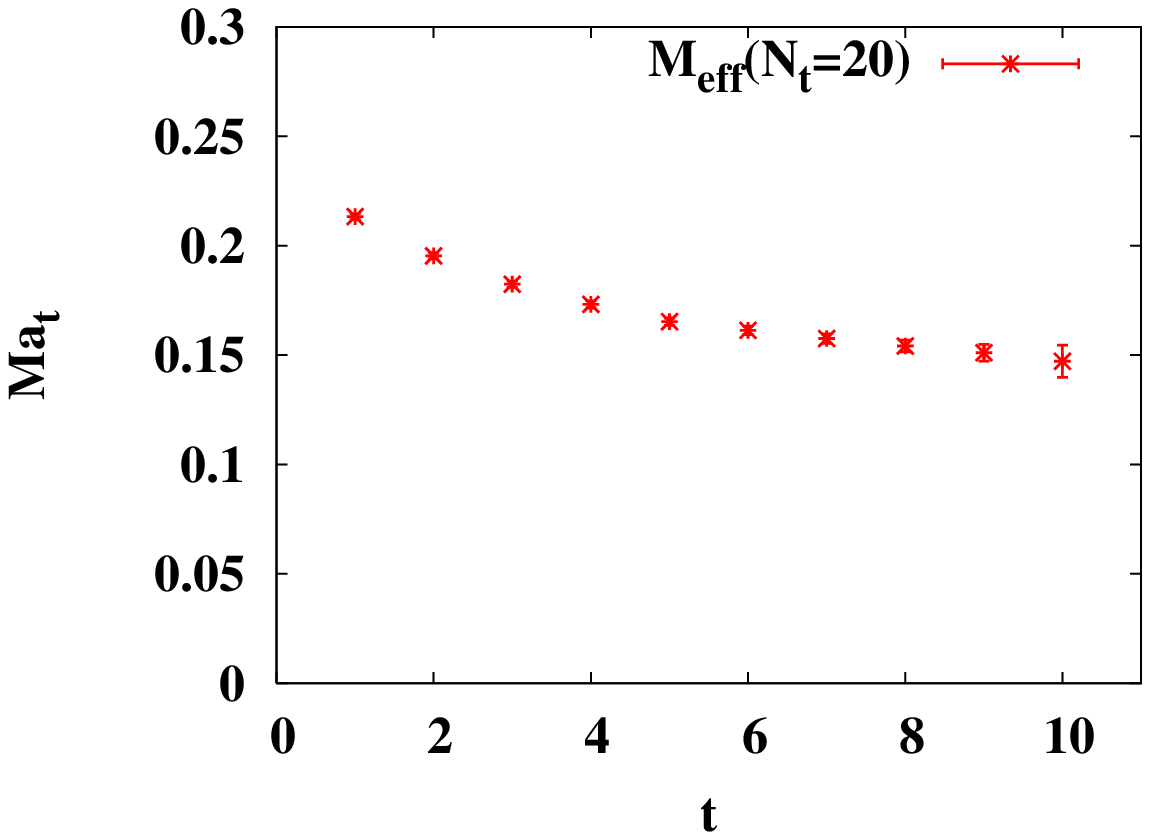}} \\
    \end{tabular}
    \caption{Similar to Fig.~\ref{opt01}, but in $E^{++}$ channel.}
    \label{opt09}
  \end{center}
\end{figure}

\begin{figure}[thb]
  \begin{center}
    \begin{tabular}{cc}
      \resizebox{40mm}{!}{\includegraphics{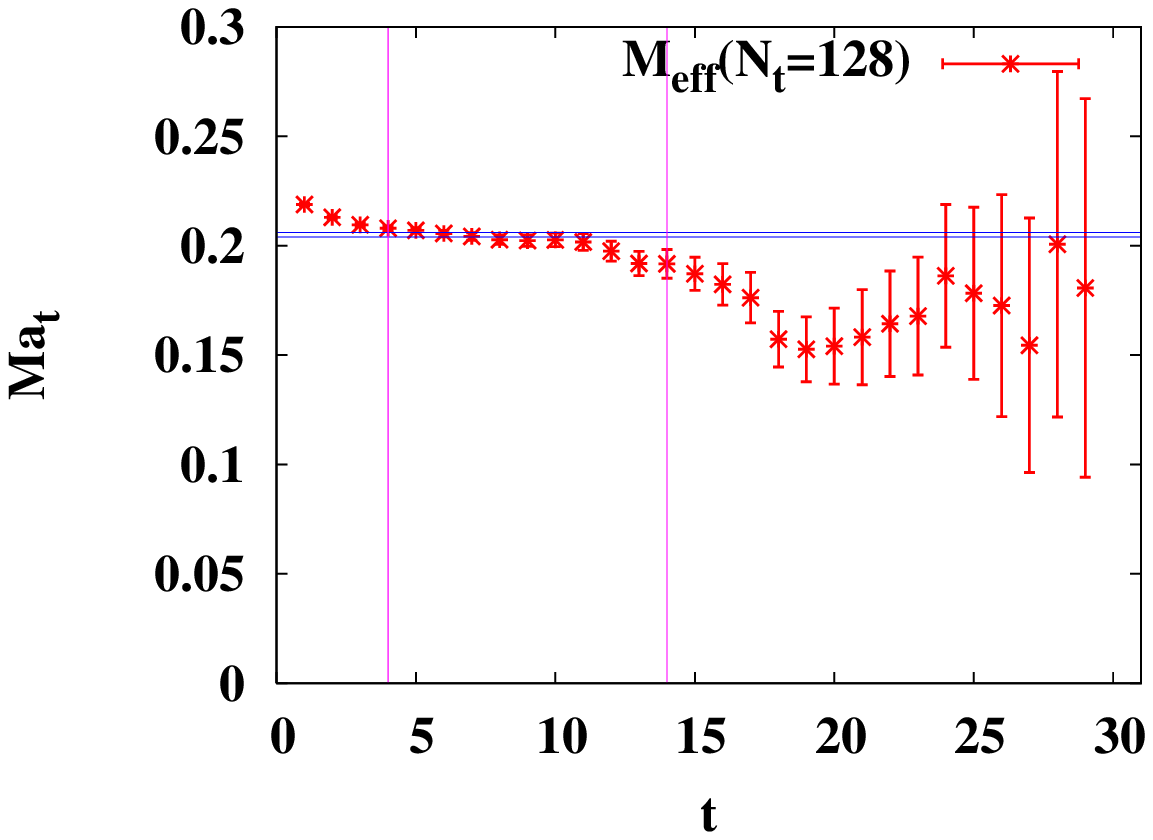}} &
      \resizebox{40mm}{!}{\includegraphics{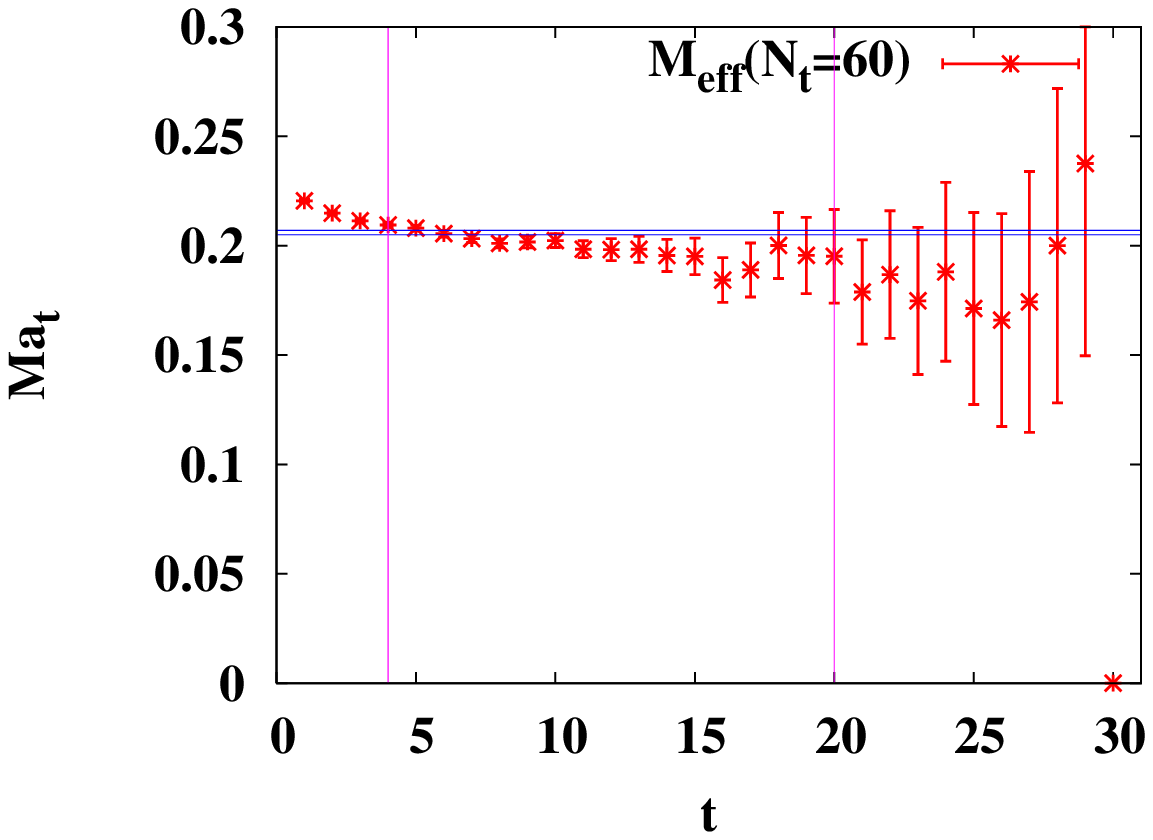}} \\
      \resizebox{40mm}{!}{\includegraphics{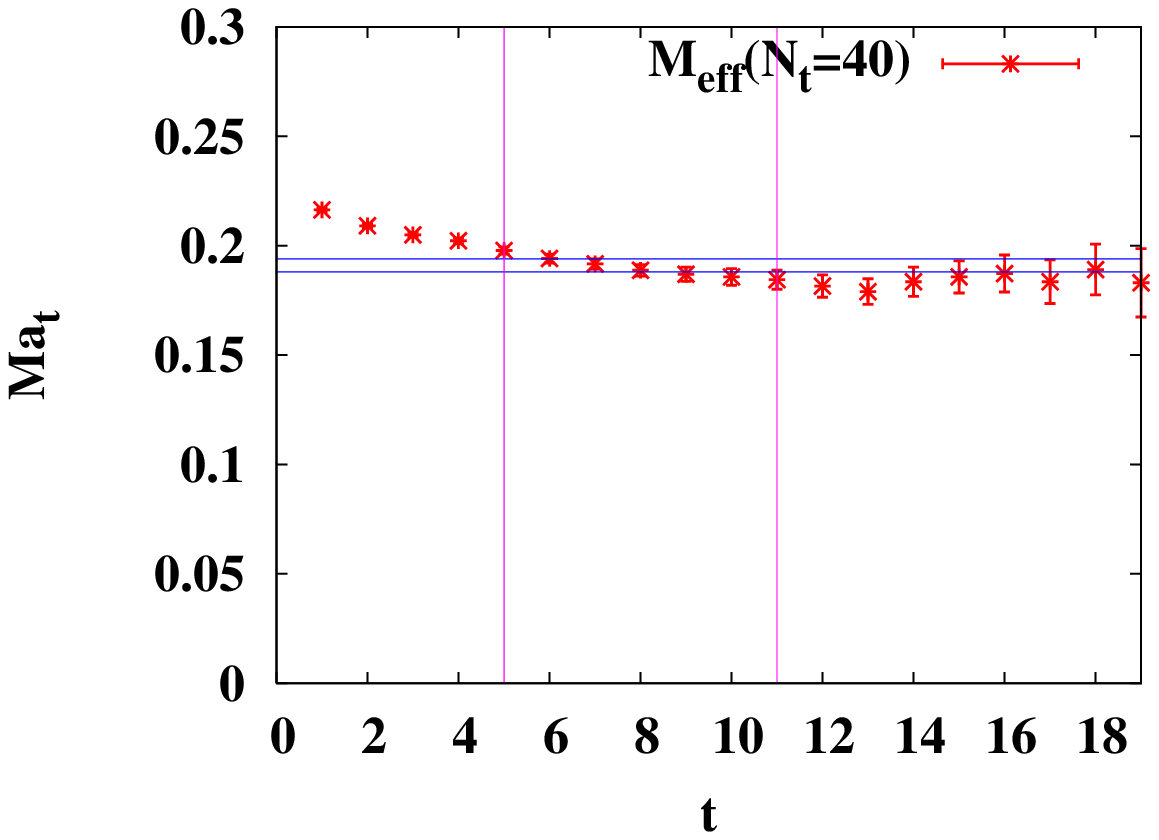}} &
      \resizebox{40mm}{!}{\includegraphics{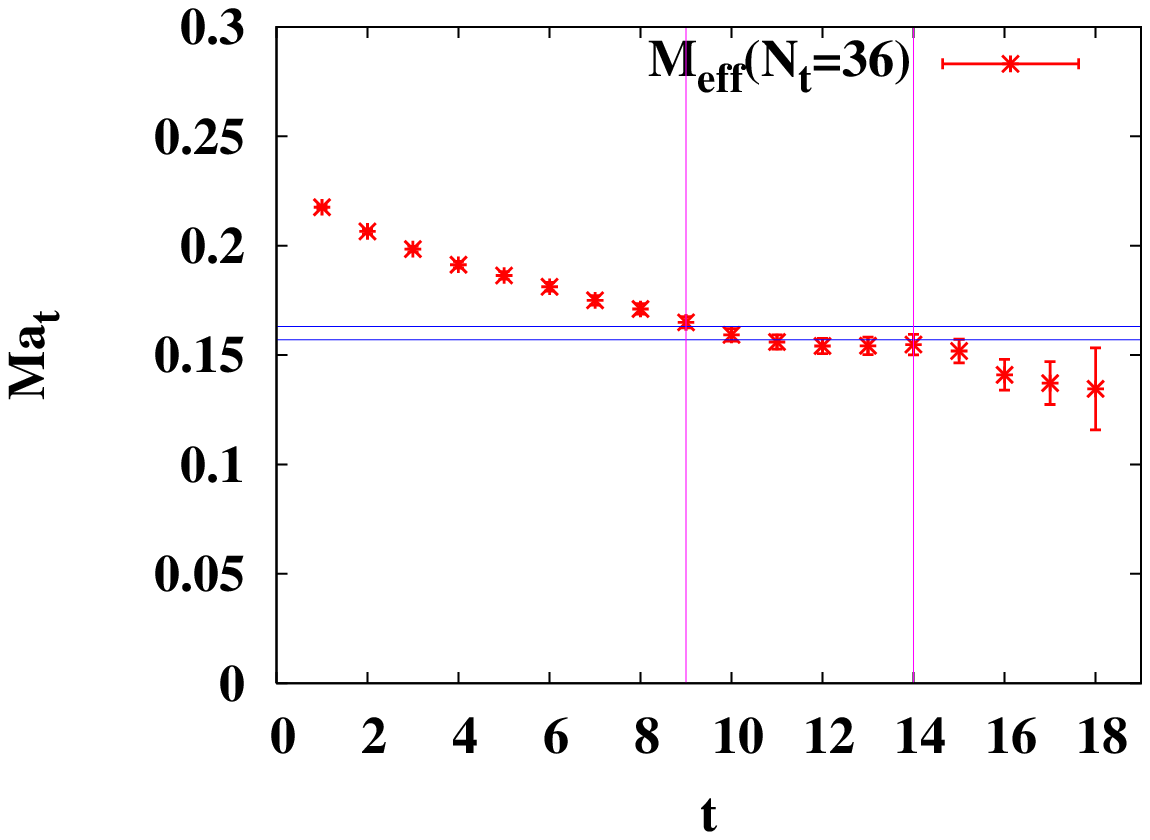}} \\
      \resizebox{40mm}{!}{\includegraphics{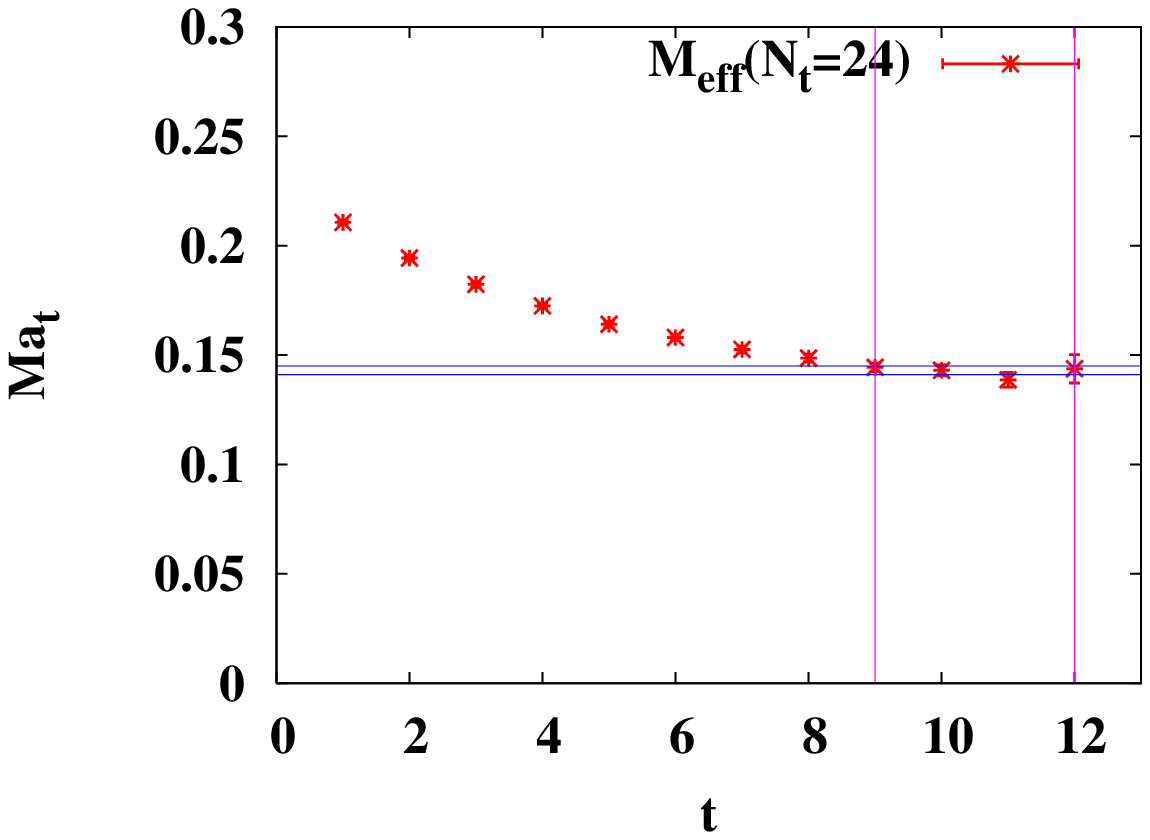}} &
      \resizebox{40mm}{!}{\includegraphics{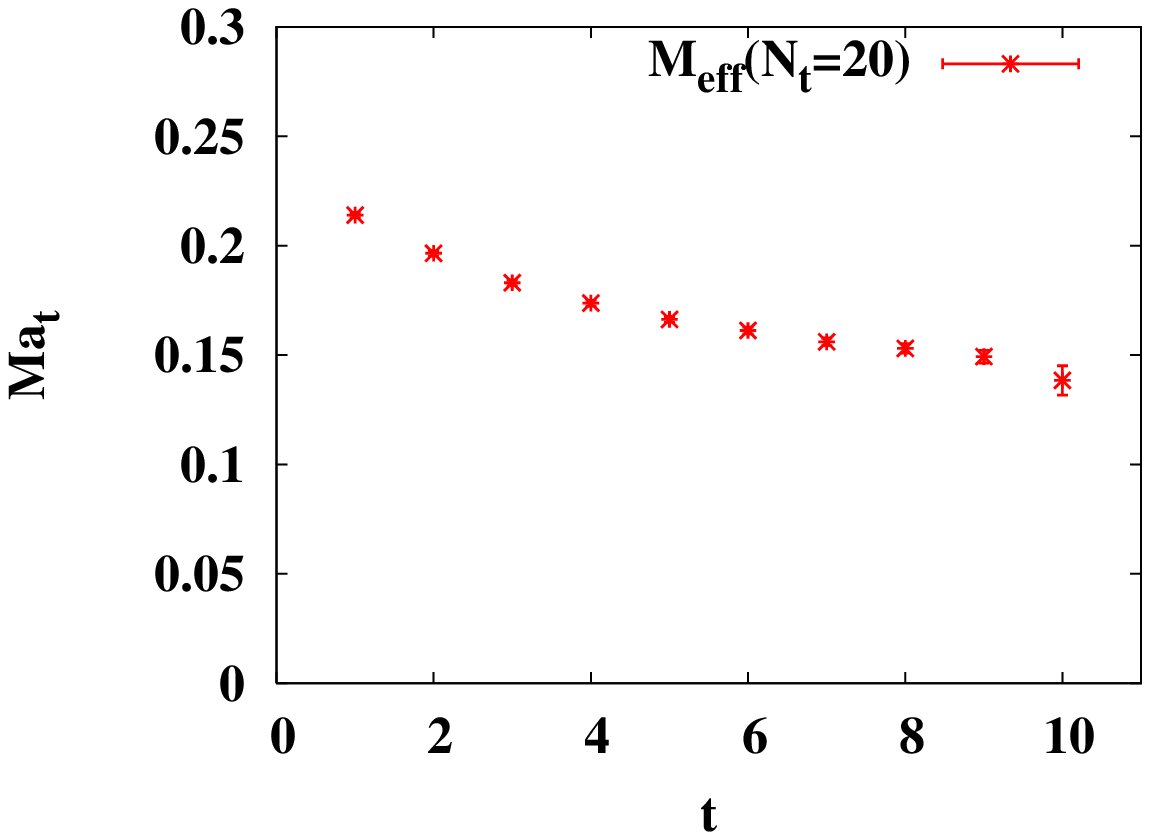}} \\
    \end{tabular}
    \caption{Similar to Fig.~\ref{opt01}, but in $T_{2}^{++}$ channel.}
    \label{opt17}
  \end{center}
\end{figure}

In Fig.~\ref{opt01}, Fig.~\ref{opt03}, Fig.~\ref{opt09}, and
Fig.~\ref{opt17} are shown the effective masses with jackknife
errors at various temperatures in $A_1^{++}$, $A_1^{-+}$, $E^{++}$,
and $T_2^{++}$ channels, respectively. The vertical lines indicate
the time window $[t_1, t_2]$ over which the single-cosh fittings are
carried out, while the horizontal lines illustrate the best-fit
result of pole masses (in each figure panel the double horizontal
lines give the error band estimated by jackknife analysis). These
figures exhibit some common features: At the temperatures below
$T_c$ ($N_t=$ 128, 80, 40), the effective mass plateaus show up
almost from right the beginning of $t$, as it should be for the
optimal glueball operators, while at $T>T_c$ ($N_t=$ 36, 24, and
20), the plateaus appear later and later in time, and even do not
exist at $N_t=20$ ($T=1.90T_c$). This observation can be interpreted
as follows. Since the effective masses are calculated based on
Eq.~\ref{cosh_function}, the very early appearance of the plateaus
below $T_c$ implies that the thermal correlators $C(t,T)$ of the
optimal operators are surely dominated by the ground state and can
be well described by the function form of Eq.~\ref{cosh_function}.
In other words, the picture of weakly interacting glueball-like
modes makes sense for the state of matter below $T_c$. While at
$T>T_c$, the later appearance and the narrower size plateaus signal
that the picture of the state of matter is distinct from that at
$T<T_c$. However, because of the existence of effective mass
plateaus up to $T\sim 1.58T_c$($N_t=24$), the possibility that
glueball-like modes survive at this high temperature cannot be
excluded.
\begin{table*}
\caption{The pole masses (in units of $a_t^{-1}$) in all the 20
$R^{PC}$ channels are extracted at all the
temperatures.\label{glueball}}
\begin{ruledtabular}
\begin{tabular}{cccccccccccc}
 $R^{PC}$   &   128 &   80  &   60  &   48  &   44  &   40  &   36  &   32  &  28 &   24  \\
  \hline
$A_{1}^{++}$  & 0.140( 2)& 0.144( 3)& 0.144( 2)& 0.143( 3)& 0.140(2)
              & 0.140( 3)& 0.132( 4)& 0.126( 2)& 0.122( 4)& 0.116( 3)\\
$A_{1}^{+-}$  & 0.441( 3)& 0.435( 3)& 0.434( 5)& 0.437( 4)& 0.432(
4)
              & 0.435( 5)& 0.399( 6)& 0.322( 9)& 0.267(16)& 0.241(13)\\
$A_{1}^{-+}$  & 0.221( 3)& 0.225( 2)& 0.222( 2)& 0.225( 2)& 0.218(
3)
              & 0.222( 2)& 0.183( 5)& 0.174( 3)& 0.155( 4)& 0.146( 4)\\
$A_{1}^{--}$  & 0.475( 6)& 0.453( 8)& 0.447( 9)& 0.464( 7)& 0.473(
6)
              & 0.468( 6)& 0.426(12)& 0.417(10)& 0.287(19)& 0.253(18)\\
&&&&&&&&&&\\
 $A_{2}^{++}$  & 0.323( 4)& 0.327( 4)& 0.326( 4)&0.330( 2)& 0.326( 4)
              & 0.332( 3)& 0.282( 7)& 0.249( 8)& 0.224( 9)& 0.208( 9)\\
$A_{2}^{+-}$  & 0.302( 5)& 0.308( 3)& 0.308( 5)& 0.312( 3)& 0.312(
5)
              & 0.308( 6)& 0.268( 6)& 0.241( 7)& 0.220( 8)& 0.201( 6)\\
$A_{2}^{-+}$  & 0.450( 5)& 0.449( 7)& 0.446( 5)& 0.440( 6)& 0.452(
4)
              & 0.448( 5)& 0.396(10)& 0.340(11)& 0.330(12)& 0.250(14)\\
$A_{2}^{--}$  & 0.387( 3)& 0.388( 3)& 0.385( 4)& 0.390( 5)& 0.376(
4)
              & 0.375( 4)& 0.354( 7)& 0.293( 7)& 0.268(10)& 0.214( 9)\\
&&&&&&&&&&\\
$E^{++}$      & 0.210( 1)& 0.205( 1)& 0.207( 1)& 0.209(2)&0.206(1)
              & 0.189( 4)& 0.167( 4)& 0.153( 3)& 0.143( 3)& 0.139( 2)\\
$E^{+-}$      & 0.401( 2)& 0.403( 2)& 0.401( 2)& 0.394( 4)& 0.400(
2)
              & 0.395( 3)& 0.375( 4)& 0.311( 6)& 0.261( 7)& 0.230( 7)\\
$E^{-+}$      & 0.273( 1)& 0.266( 1)& 0.264( 2)& 0.273( 2)& 0.275(
1)
              & 0.262( 2)& 0.218( 4)& 0.196( 4)& 0.183( 4)& 0.181( 4)\\
$E^{--}$      & 0.374( 1)& 0.368( 2)& 0.360( 2)& 0.361( 3)& 0.363(
3)
              & 0.352( 4)& 0.308( 8)& 0.262( 6)& 0.231( 6)& 0.213( 6)\\
&&&&&&&&&&\\
$T_{1}^{++}$  & 0.327( 2)& 0.326( 4)& 0.327( 2)& 0.334( 2)& 0.331(
2)
              & 0.312( 5)& 0.287( 7)& 0.266( 3)& 0.227( 6)& 0.215( 4)\\
$T_{1}^{+-}$  & 0.278( 1)& 0.274( 2)& 0.265( 3)& 0.278( 2)& 0.281(
1)
              & 0.261( 3)& 0.207( 6)& 0.199( 2)& 0.181( 4)& 0.175( 2)\\
$T_{1}^{-+}$  & 0.372( 2)& 0.377( 4)& 0.371( 3)& 0.380( 2)& 0.374(
2)
              & 0.370( 3)& 0.331( 5)& 0.289( 7)& 0.248( 7)& 0.230( 5)\\
$T_{1}^{--}$  & 0.350( 4)& 0.349( 2)& 0.344( 3)& 0.351( 2)& 0.350(
2)
              & 0.343( 3)& 0.272( 8)& 0.252( 5)& 0.212( 6)& 0.201( 5)\\
&&&&&&&&&&\\
$T_{2}^{++}$  & 0.205( 1)& 0.209( 1)& 0.206( 1)& 0.205( 1)& 0.207(
2)
              & 0.191( 3)& 0.160( 3)& 0.152( 2)& 0.148( 2)& 0.143( 2)\\
$T_{1}^{+-}$  & 0.322( 2)& 0.317( 2)& 0.310( 4)& 0.317( 3)& 0.320(
2)
              & 0.303( 5)& 0.276( 5)& 0.250( 3)& 0.201( 4)& 0.190( 4)\\
$T_{1}^{-+}$  & 0.265( 2)& 0.260( 3)& 0.264( 2)& 0.273( 3)& 0.272(
2)
              & 0.264( 2)& 0.240( 3)& 0.213( 3)& 0.187( 4)& 0.183( 4)\\
$T_{1}^{--}$  & 0.368( 2)& 0.358( 3)& 0.364( 3)& 0.358( 4)& 0.367(
2)
              & 0.353( 5)& 0.282(13)& 0.254( 6)& 0.235( 6)& 0.220( 4)\\
\end{tabular}
\end{ruledtabular}
\end{table*}

\begin{figure}
\includegraphics[height=5cm]{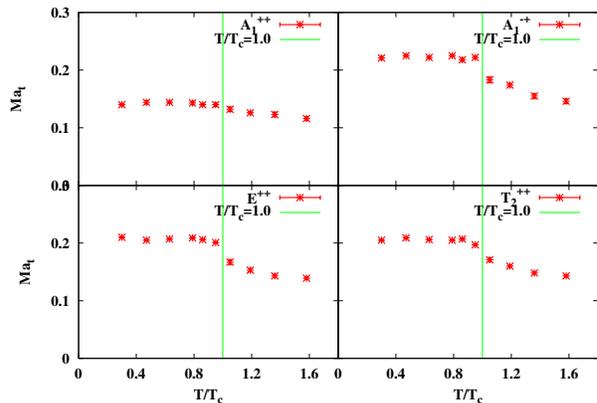}
\caption{The $T$-dependence of pole masses $A_{1}^{++}$,
$A_{1}^{-+}$ $E^{++}$, and $T_{2}^{++}$ glueballs.
\label{opt_total}}
\end{figure}

The pole masses in all 20 $R^{PC}$ channels are extracted in units
$a_t^{-1}$ at all temperatures and are shown in
Table~\ref{glueball}. Specifically, with the lattice spacing
determined in Sec. II, the pole masses of $A_{1}^{++}$,
$A_{1}^{-+}$, $E^{++}$ and $T_{2}^{++}$ at $T\simeq 0$ in physical
units are $M_{A_{1}^{++}}$=1.576(22)GeV,
$M_{A_{1}^{-+}}$=2.488(34)GeV, $M_{E^{++}}\simeq
M_{T_{2}^{++}}$=2.364(11)GeV, respectively, which are in agreement
with that of previous
studies~\cite{prd56,prd60,prd73,npb221,plb309,npb314,prl75,outp}.
From the table, one can see that the behaviors of the pole masses
with respect to the temperature in all 20 channels are uniform: the
pole masses keep almost constant with the temperature increasing
from $0.30T_c$ to right below $T_c$ ($0.95 T_c$), and start to
reduce gradually when $T>T_c$. When $T$ increases up to $1.90T_c$,
the pole masses cannot be extract reliably through the single-cosh
fit for the lack of clear effective mass plateaus.
Figure~\ref{opt_total} illustrates these behavior of pole masses in
$A_{1}^{++}$, $A_{1}^{-+}$, $E^{++}$ and $T_{2}^{++}$ channels.

These results imply that glueballs can be very stable below $T_c$
and survive up to $1.6T_c$. This coincides with the thermal
properties of heavy quarkonia observed by model calculation and
lattice numerical studies
\cite{soft_modes,quark_number,gupta,prl92,prd69,prd74,jhw2005,npa783,prd63,ijmpa16},
but different from the observation of a previous lattice study on
glueballs where the observed pole-mass reduction start even at
$T\simeq 0.8 T_{c}$~\cite{prd66}.

\subsection{Breit-Wigner analysis}
In the single-cosh analysis, it is seen that, when the temperature
increases up to $T_c$, the thermal correlators can be well described
by Eq.~\ref{cosh_function} and the pole masses of glueballs are
insensitive to $T$. This is in agreement with the picture that the
state of matter below $T_c$ are made up of weakly interacting
glueball-like modes. When $T>T_c$, the thermal correlators deviate
from Eq.~\ref{cosh_function} more and more. This observation implies
that the degrees of freedom are very different from that when
$T<T_c$. Theoretically in the deconfined phase, gluons can be
liberated from hadrons. However, the study of the equation of state
shows that the state of the matter right above $T_c$ is far from a
perturbative gluon gas. In other words, the gluons in the
intermediate temperature above $T_c$ may interact strongly with each
other and glueball-like resonances can possibly be formed. Thus
different from bound states at low temperature, thermal glueballs
can acquire thermal width due to the thermal scattering between
strongly interacting gluons and the magnitudes of the thermal widths
can signal the strength of these types of interactions at different
temperatures.

In order to take the thermal width into consideration, we also adopt
the Breit-Wigner $\emph{Ansatz}$, which is suggested by the
pioneering work Ref.~\cite{prd66}, to analyze the thermal
correlators once more. First, we treat thermal glueballs as
resonance objects which correspond to the poles (denoted by
$\omega=\omega_0-i\Gamma$) of the retarded and advanced Green
functions in the complex $\omega\_$plane (note that conventionally
in particle physics, a resonance pole is always denoted as
$M-i\Gamma/2$ where $M$ is the mass of the resonance and $\Gamma$ is
its width.) $\omega_0$ is called the mass of the resonance glueball
and $\Gamma$ its thermal width in this work.  Secondly, we assume
that the spectral function $\rho(\omega)$ is dominated by these
resonance glueballs. Thus the spectral function is parametrized as
\begin{equation}
 \rho(\omega) =A(\delta_\Gamma(\omega-\omega_0)-\delta_\Gamma
(\omega+\omega_0)+\ldots,
\end{equation}
where $\delta_{\epsilon}$ is the Lorentzian function
\begin{equation}
\delta_\epsilon(x) = \frac{1}{\pi}{\rm Im} \left(
\frac{1}{x-i\epsilon}
\right)=\frac{1}{\pi}\frac{\epsilon}{x^2+\epsilon^2},
\end{equation}
and "$\ldots$" represents the terms of excited states. With this
spectral function, the thermal glueball correlator $G(t,T)$ can be
expressed as
\begin{eqnarray}
C(t,T)&=&\int\limits_{-\infty}^{\infty}\frac{d\omega}{2\pi}
\frac{\cosh(\omega(\frac{1}{2T}-t))}{2\sinh(\frac{\omega}{2T})}\nonumber\\
&\times& 2\pi A\left(
\delta_\Gamma(\omega-\omega_0)-\delta_\Gamma(\omega+\omega_0)\right)+\ldots.
\end{eqnarray}
\begin{figure*}[htb]
  \begin{center}
    \begin{tabular}{ccc}
     (a)~$N_t=128$($T/T_c=0.32$) & (b)~$N_t=36$($T/T_c=1.09$) &(c)~$N_t=20$($T/T_c=1.97$)\\
      \resizebox{50mm}{!}{\includegraphics{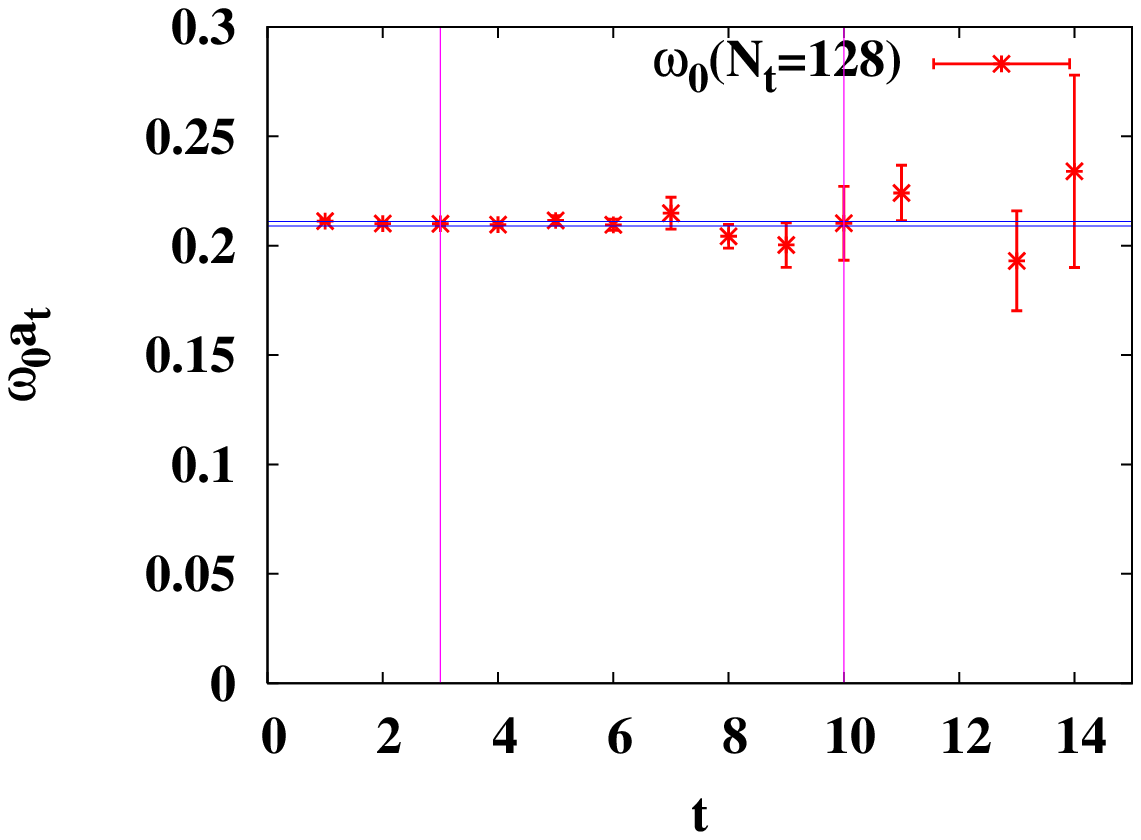}} &
      \resizebox{50mm}{!}{\includegraphics{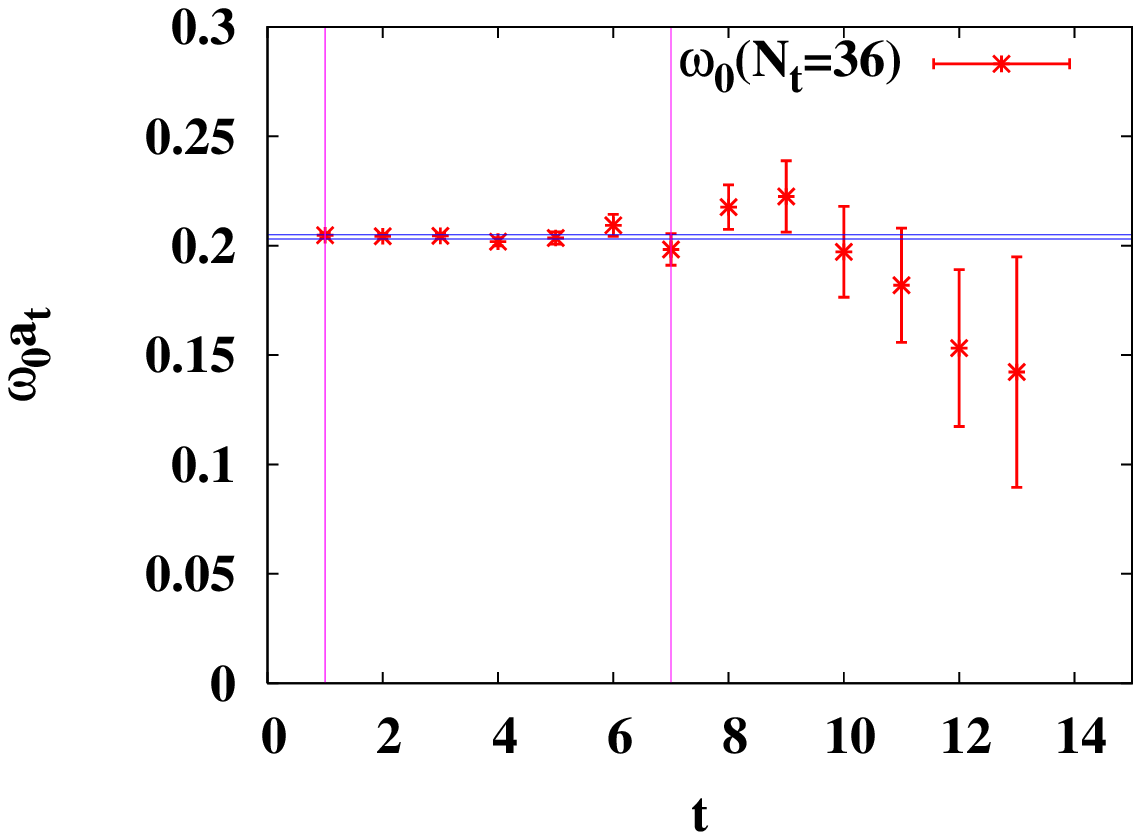}} &
      \resizebox{50mm}{!}{\includegraphics{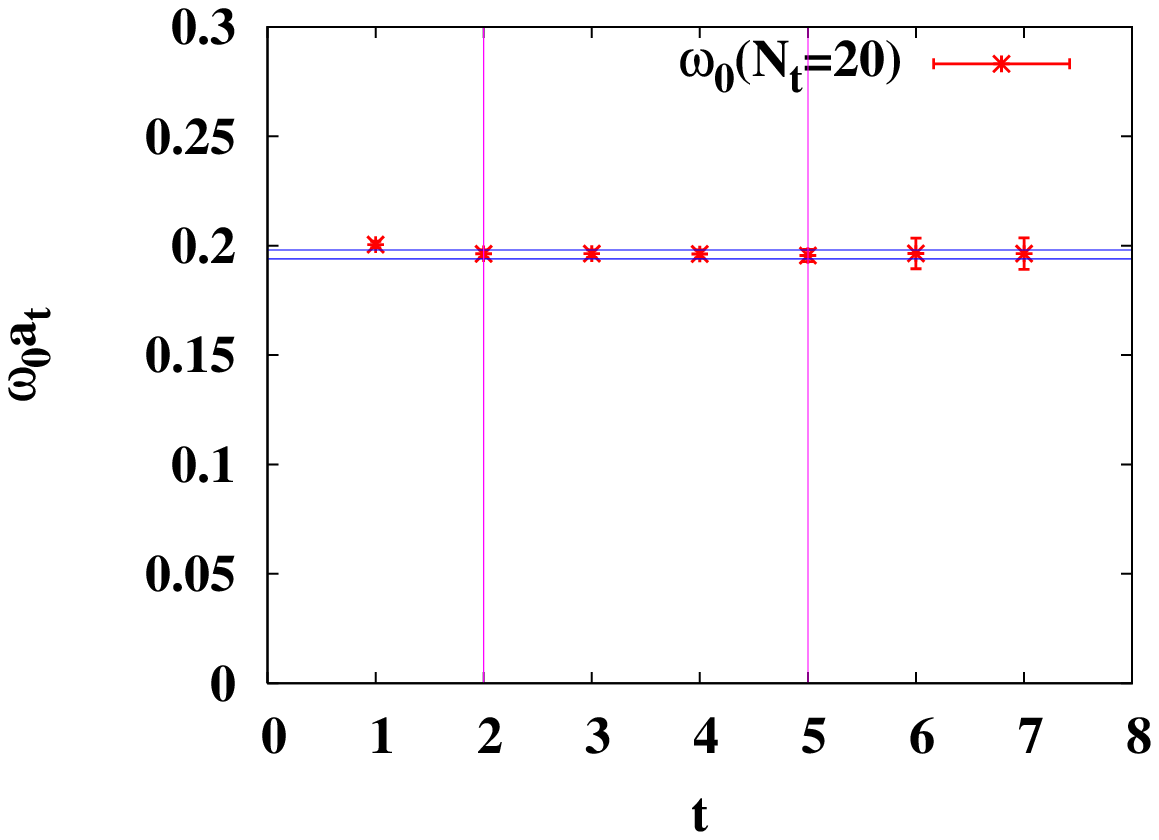}} \\
      \resizebox{50mm}{!}{\includegraphics{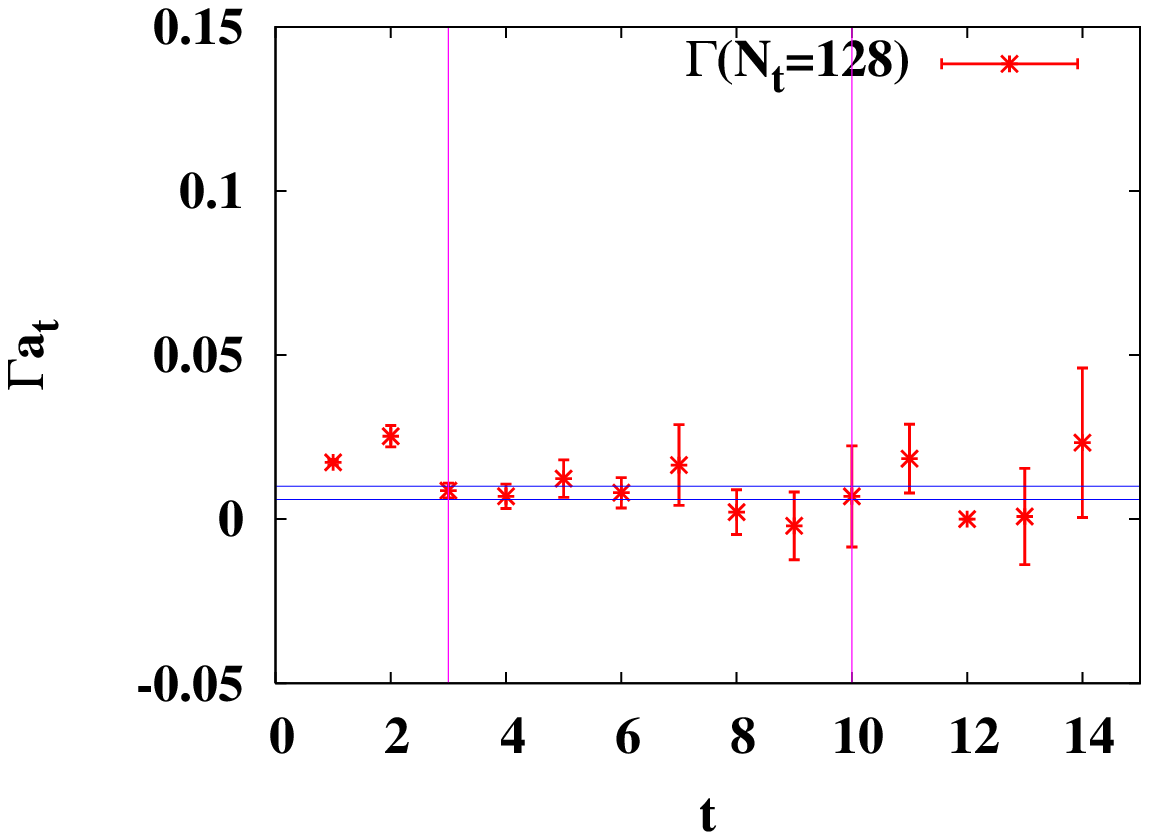}} &
      \resizebox{50mm}{!}{\includegraphics{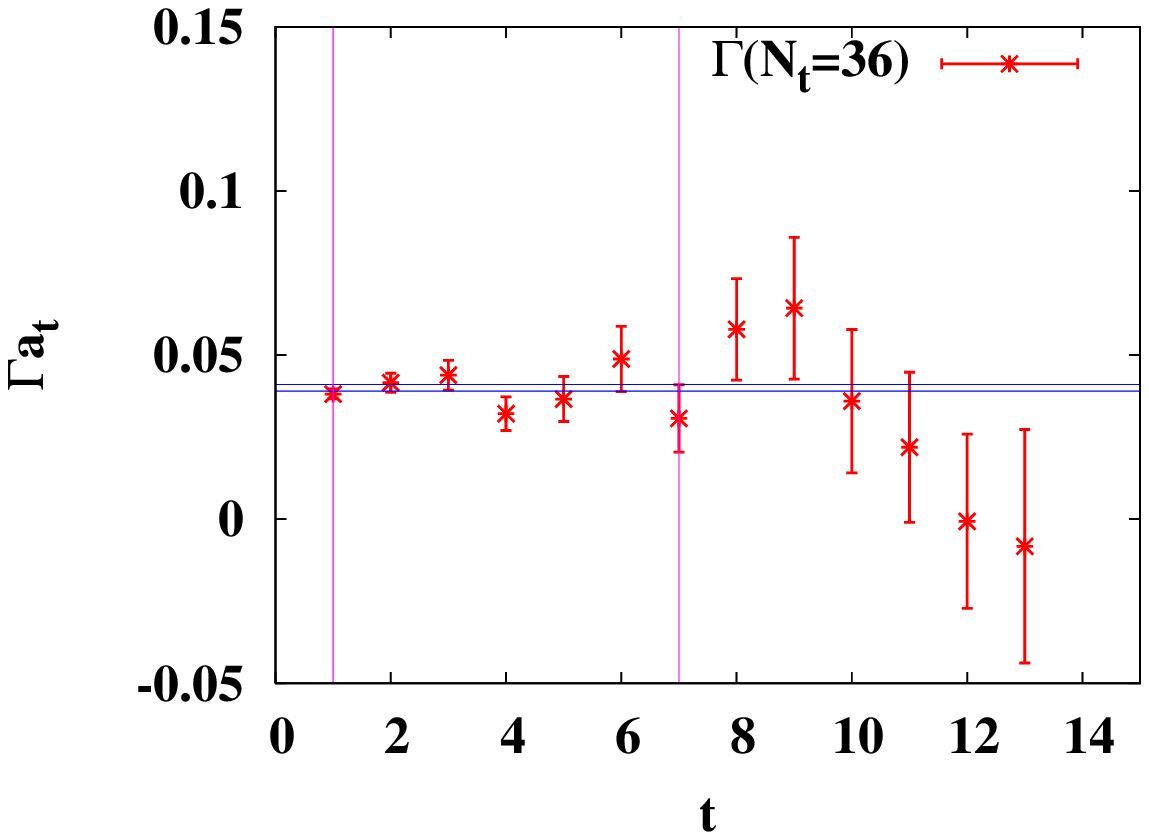}}  &
      \resizebox{50mm}{!}{\includegraphics{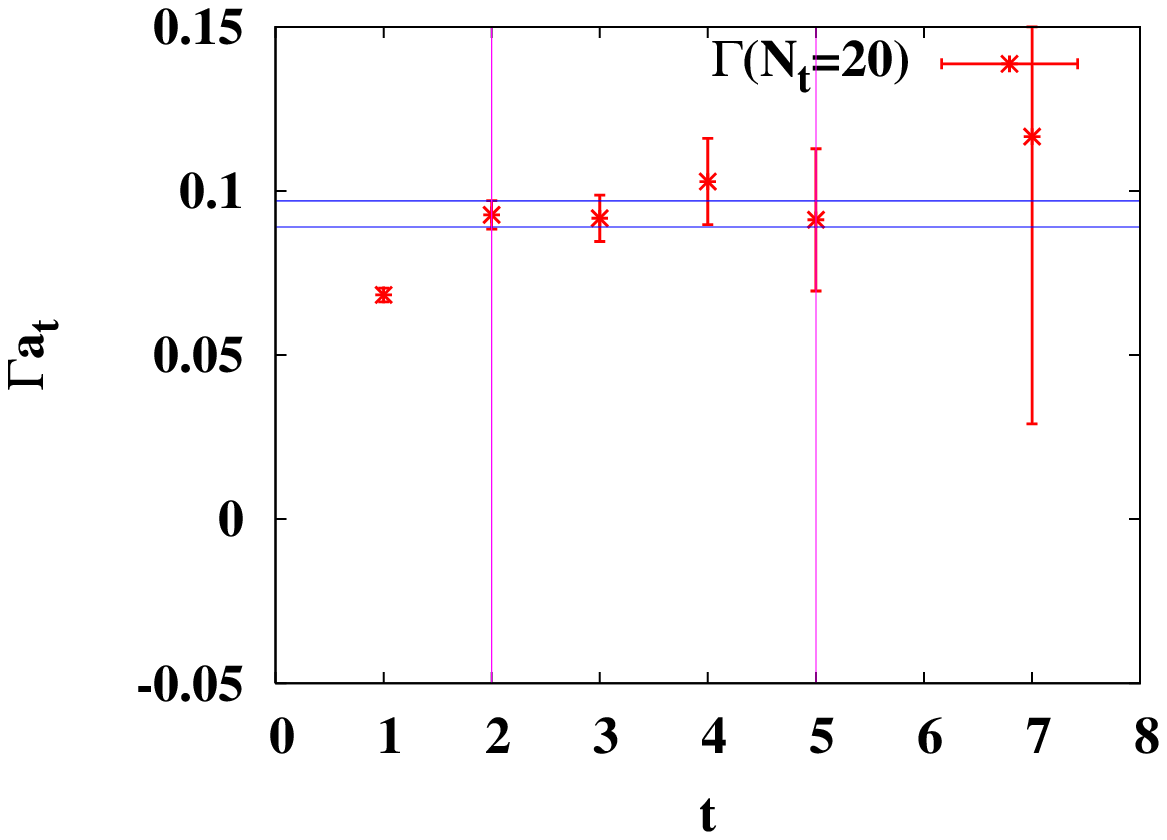}}
    \end{tabular}
    \caption{\label{opt17_bw} Determinations of the fit range $[t_1,t_2]$
          in $T_2^{++}$ channel at $N_t=$ 128, 36, and 20. In each
          row, $\omega_0^{\rm (eff)}(t)$ and
          $\Gamma^{\rm (eff)}(t)$ obtained by solving Eq.~\ref{solution}
          are plotted by data points with jackknife error bars.
          $[t_1,t_2]$ are chosen to include the time slices
          between the two vertical lines, where $\omega_0^{\rm (eff)}(t)$ and
          $\Gamma^{\rm (eff)}(t)$ show up plateaus
          simultaneously. The best-fit results of $\omega_0$ and
          $\Gamma$ through the function $g_\Gamma(t)$ are
          illustrated by the horizontal lines.}

  \end{center}
\end{figure*}
The integral on the right hand side of above equation, denoted by
$g_\Gamma (t)$, can be calculated explicitly as
\begin{widetext}
\begin{equation}
\label{fit_fun}
 g_\Gamma(t)=A\left[ {\rm Re} \left(
\frac{\cosh((\omega_0+i\Gamma)(\frac{1}{2T}-t))}
{\sinh(\frac{(\omega_0+i\Gamma)}{2T})} \right)+ 2\omega_0 T
\sum\limits_{n=1}^{\infty}\cos\left(2\pi nTt \right) \left\{
\frac{1}{(2\pi nT+\Gamma)^2+\omega_0^2}-(n\rightarrow -n)
 \right\}
 \right],
\end{equation}
\end{widetext}
which can be used as the fit function to extract $\omega_0$ and
$\Gamma$ from the thermal correlators obtained from the numerical
calculation. Practically, the infinite series in the above equation
is truncated by setting the upper limit of the summation to be 50,
which is tested to be enough for all the cases considered in this
work.

In the present study, we carry out the Breit-Wigner analysis in
$A_1^{++}$, $A_1^{-+}$, $E^{++}$, and $T_2^{++}$ channels, whose
continuum correspondences are $0^{++}$, $0^{-+}$, and $2^{++}$.
Although the variational method is exploited to enhance the
contribution of the ground state to the thermal correlators, the
contributions from higher spectral components cannot be eliminated
completely. Therefore, the fit range must be chosen properly where
the contribution of the ground state dominates. We take the strategy
advocated in Ref.\cite{prd66} as follows. For a given correlator
$C(t,T)$, the effective peak position $\omega_0^{\rm (eff)}(t)$ and
the effective width $\Gamma^{\rm (eff)}(t)$ are obtained by solving
the equations
\begin{eqnarray}
\label{solution}
\frac{g_\Gamma(t)}{g_\Gamma(t+1)}&=& \frac{C(t,T)}{C(t+1,T)},\nonumber\\
\frac{g_\Gamma(t+1)}{g_\Gamma(t+2)}&=& \frac{C(t+1,T)}{C(t+2,T)}.
\end{eqnarray}
The statistical errors of $\omega_0^{\rm (eff)}(t)$ and $\Gamma^{\rm
(eff)}(t)$ can be estimated through the jackknife analysis. Thus the
fit range, denoted by $[t_1, t_2]$, is chosen to be the time range
where $\omega_0^{\rm (eff)}(t)$ and $\Gamma^{\rm (eff)}(t)$ show up
plateaus simultaneously. For example, the procedure in $T_2^{++}$
channel is illustrated in Fig.~\ref{opt17_bw} for $N_t = 128, 36,
20$ (corresponding to the temperature $T/T_c=0.30, 1.05, 1.90$),
where the fit ranges $[t_1,t_2]$ are determined to include the time
slices between the two vertical lines in each figure.

\begin{figure}[ht]
\includegraphics[height=5cm]{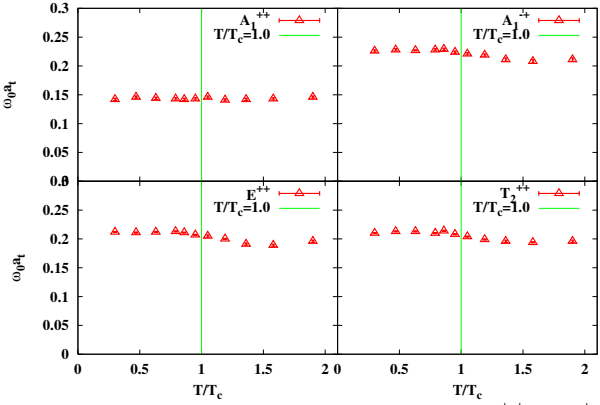}
\caption{$\omega_0$'s are plotted versus $T/T_c$ for $A_{1}^{++}$,
$A_{1}^{-+}$ $E^{++}$, and $T_{2}^{++}$ channels. The vertical lines
indicate the critical temperature.
       \label{opt_total_om}}
\end{figure}

\begin{figure}[ht]
\includegraphics[height=5cm]{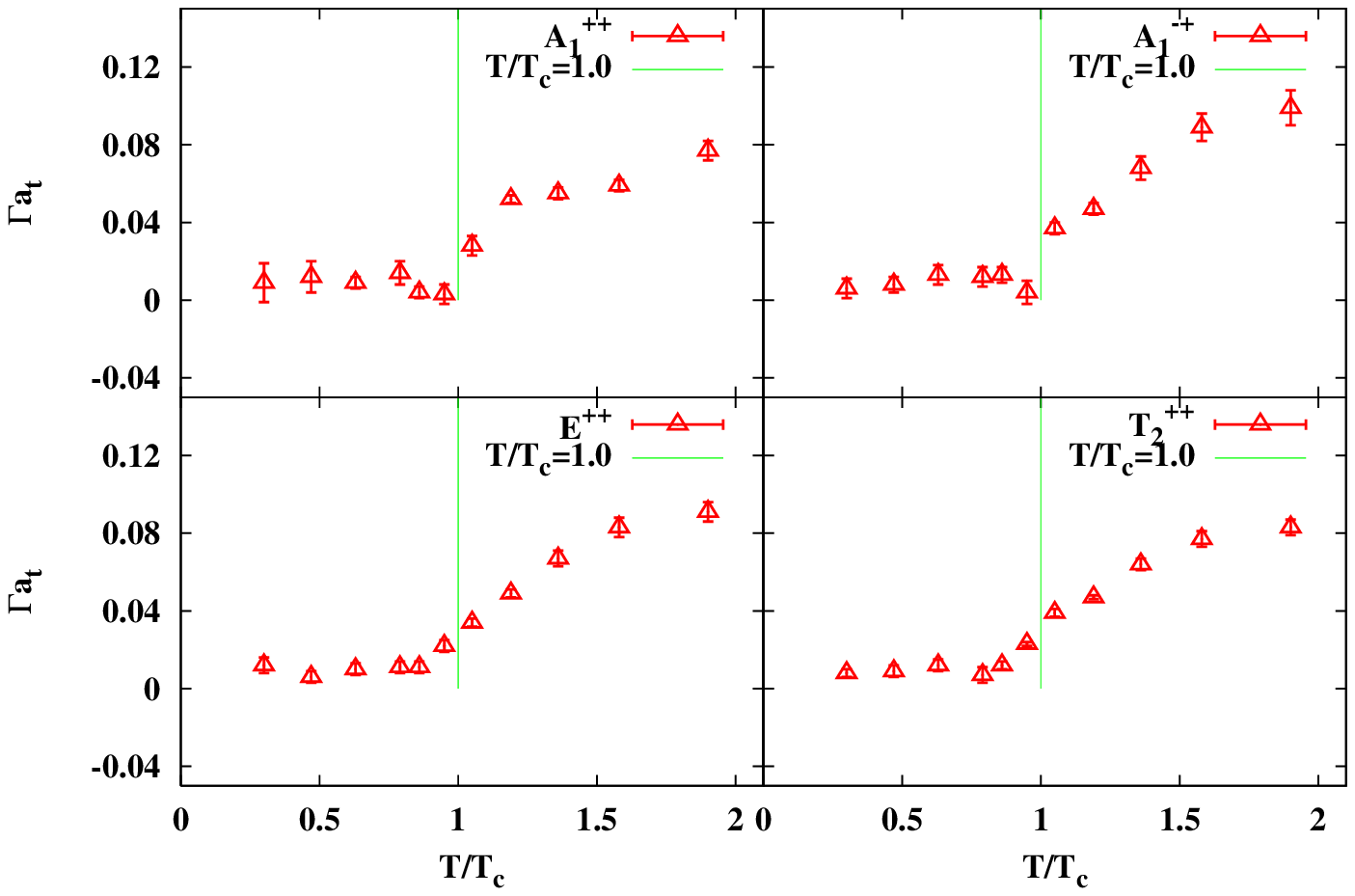}
\caption{$\Gamma$'s are plotted versus $T/T_c$ for $A_{1}^{++}$,
$A_{1}^{-+}$ $E^{++}$, and $T_{2}^{++}$ channels. The vertical lines
indicate the critical temperature.
       \label{opt_total_gm}}
\end{figure}
\begin{figure*}[ht]
  \begin{center}
    \begin{tabular}{ccc}
         (a) $A_1^{-+}$   & (b) $E^{++}$ & (c) $T_2^{++}$\\
      \resizebox{50mm}{!}{\includegraphics{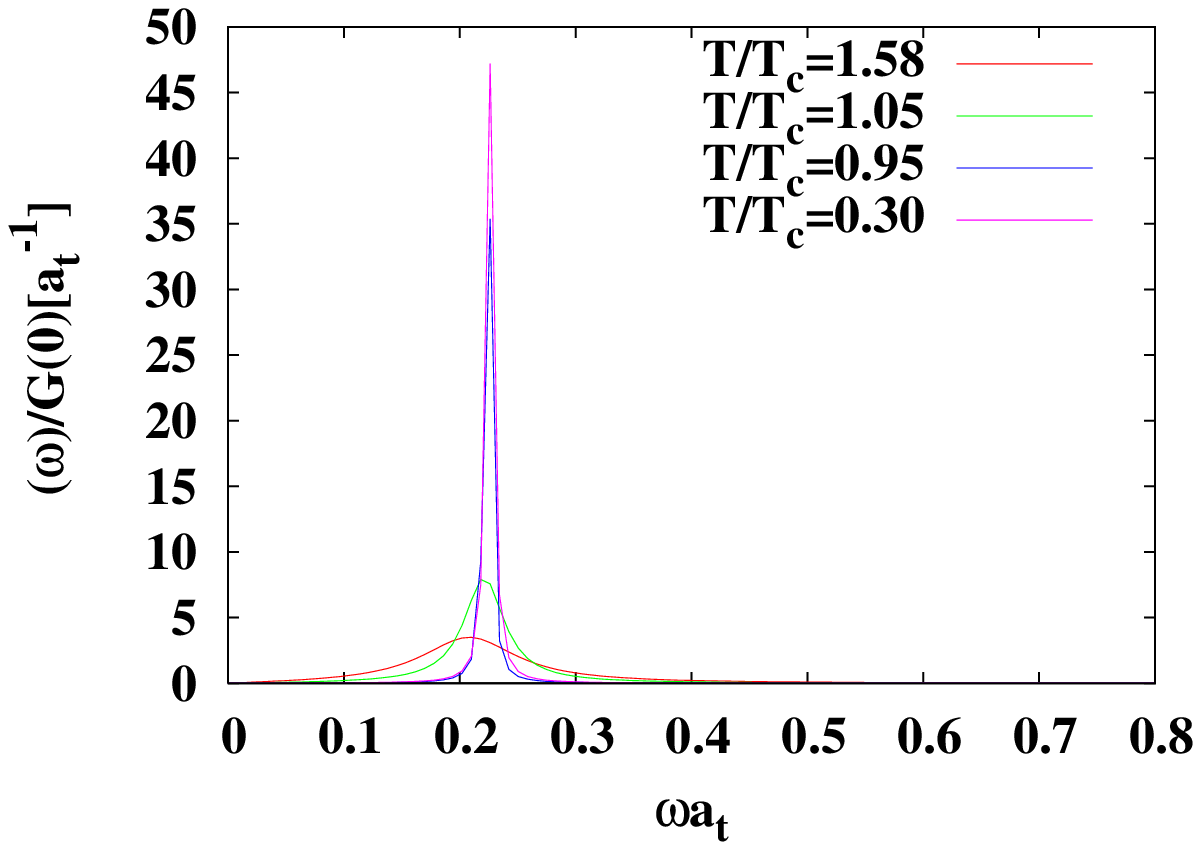}} &
      \resizebox{50mm}{!}{\includegraphics{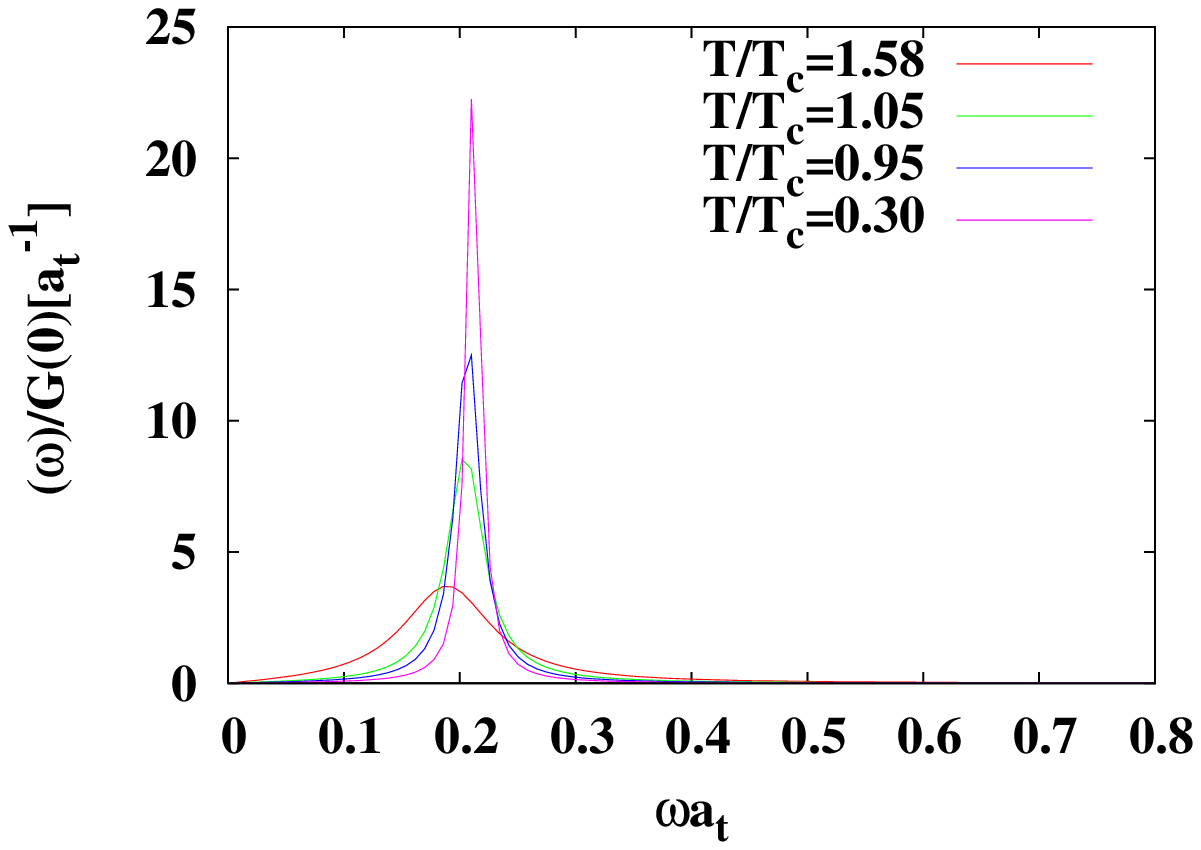}} &
      \resizebox{50mm}{!}{\includegraphics{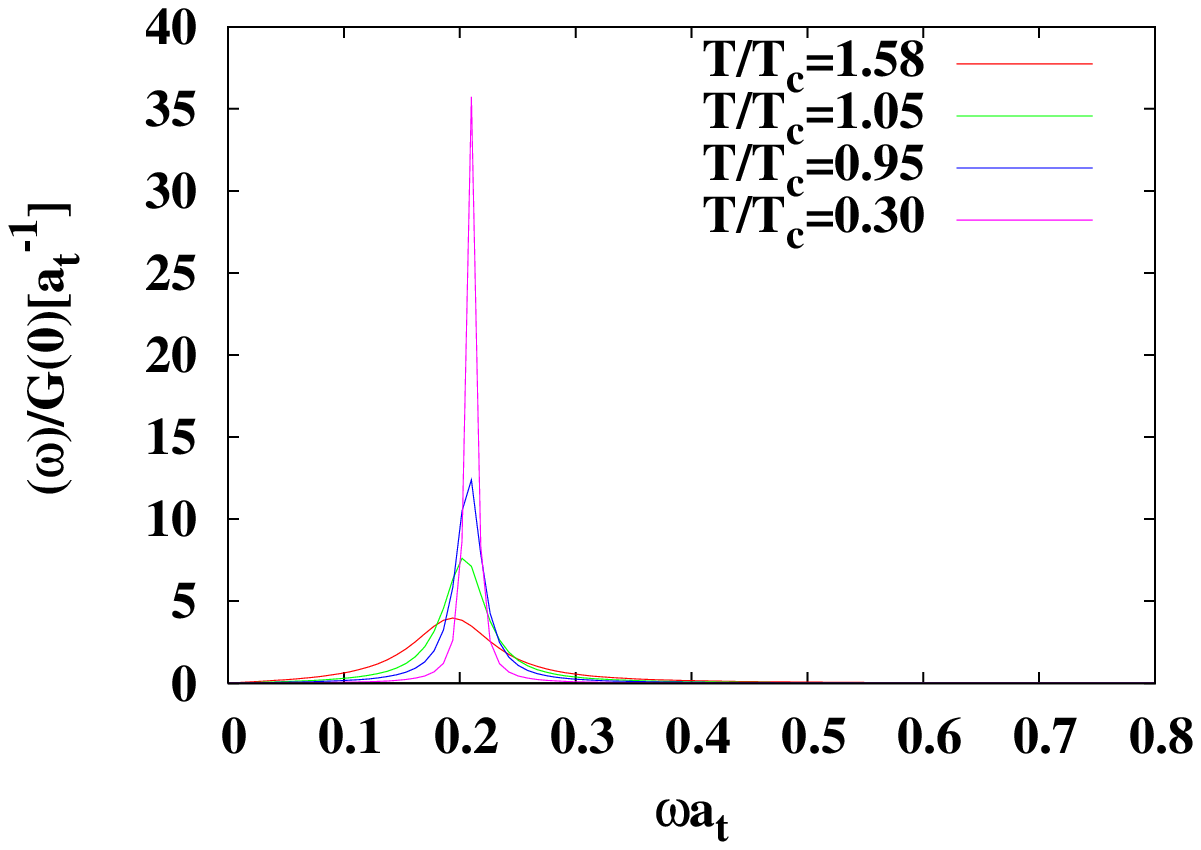}}
    \end{tabular}
    \caption{Plotted are the spectral function $\rho(\omega)$ at $T/T_c=0.30$, 0.95,1.05 and
    1.58
    with the best-fit parameters. Panel (a)$\_$(c) are for
    $A_1^{-+}$, $E^{++}$, and $T_2^{++}$ channels, respectively.
    \label{peak}}
  \end{center}
\end{figure*}
\begin{table}[tb]
\caption{ The best-fit $\omega_0$ and $\Gamma$ of $A_1^{++}$ channel
at different $T$ through the Breit-Wigner fit. Also listed are the
fit window $[t_1, t_2]$ and the chi-square per degree of freedom,
$\chi^2/d.o.f$. \label{bw_opt01}}
\begin{ruledtabular}
\begin{tabular}{cccccc}
  $N_t$ & $T/T_{c}$ & $\omega_{0}$ & $\Gamma$ & $[t_1,t_2]$   &  $\chi^{2}/d.o.f$  \\
  \hline
 128 &  0.30   &  0.142(2) &  0.008(10) & (3, 6) &  1.266\\
  80 &  0.47   &  0.146(2) &  0.013(9) & (2, 4) &  0.921\\
  60 &  0.63   &  0.144(2) &  0.008(3) & (2, 7) &  0.135\\
  48 &  0.79   &  0.143(2) &  0.014(6) & (2, 4) &  0.639\\
  44 &  0.86   &  0.142(2) &  0.004(3) & (1, 7) &  0.758\\
  40 &  0.95   &  0.143(2) &  0.003(4) & (4, 7) &  0.850\\
  &&&&&\\
  36 &  1.05   &  0.146(2) &  0.028(4) & (3, 6) &  0.960\\
  32 &  1.19   &  0.141(2) &  0.053(2) & (1, 4) &  0.393\\
  28 &  1.36   &  0.142(2) &  0.056(4) & (2, 5) &  1.253\\
  24 &  1.58   &  0.143(1) &  0.059(3) & (1, 4) &  0.302\\
  20 &  1.90   &  0.146(2) &  0.077(5) & (2, 4) &  0.918
\end{tabular}
\end{ruledtabular}
\end{table}
\begin{table}[tb]
\caption{The best-fit $\omega_0$ and $\Gamma$ of $A_1^{-+}$ channel
at different $T$ through the Breit-Wigner fit. Also listed are the
fit window $[t_1, t_2]$ and the chi-square per degree of freedom,
$\chi^2/d.o.f$. \label{bw_opt03}}
\begin{ruledtabular}
\begin{tabular}{cccccc}
  $N_t$ & $T/T_{c}$ & $\omega_{0}$ & $\Gamma$ & $[t_1,t_2]$ & $\chi^{2}/d.o.f$  \\
  \hline
 128 &  0.30    &  0.226(2) &  0.006(5) & (3, 9) &  0.509  \\
  80 &  0.47    &  0.228(2) &  0.008(4) & (2, 6) &  0.640  \\
  60 &  0.63    &  0.227(1) &  0.013(5) & (2, 7) &  0.216  \\
  48 &  0.79    &  0.228(2) &  0.012(5) & (2, 6) &  0.177  \\
  44 &  0.86    &  0.229(2) &  0.013(4) & (2, 8) &  0.184  \\
  40 &  0.95    &  0.224(2) &  0.004(6) & (3, 6) &  0.549  \\
 &&&&&\\
  36 &  1.05    &  0.221(2) &  0.037(3) & (1, 8) &  0.935  \\
  32 &  1.19    &  0.219(2) &  0.047(3) & (1, 6) &  0.250  \\
  28 &  1.36    &  0.211(3) &  0.068(6) & (2, 5) &  0.091  \\
  24 &  1.58    &  0.208(3) &  0.089(7) & (2, 4) &  0.003  \\
  20 &  1.90    &  0.211(3) &  0.099(9) & (2, 6) &  0.083
\end{tabular}
\end{ruledtabular}
\end{table}
\begin{table}[tb]
\caption{The best-fit $\omega_0$ and $\Gamma$ of $E^{++}$ channel at
different $T$ through the Breit-Wigner fit. Also listed are the fit
window $[t_1, t_2]$ and the chi-square per degree of freedom,
$\chi^2/d.o.f$. \label{bw_opt09}}
\begin{ruledtabular}
\begin{tabular}{cccccc}
  $N_t$ & $T/T_{c}$ & $\omega_{0}$ & $\Gamma$ & ${t_1,t_2}$ & $\chi^{2}/DOF$  \\
  \hline
 128 &  0.30    &  0.212(1) &  0.012(4) & (2, 5) &  0.274    \\
  80 &  0.47    &  0.211(1) &  0.006(3) & (2, 8) &  0.616    \\
  60 &  0.63    &  0.212(1) &  0.010(3) & (2, 9) &  0.844    \\
  48 &  0.79    &  0.213(1) &  0.011(3) & (2, 5) &  0.206    \\
  44 &  0.86    &  0.211(1) &  0.011(3) & (2, 8) &  1.268    \\
  40 &  0.95    &  0.207(1) &  0.022(3) & (2, 6) &  0.250    \\
 &&&&&\\
  36 &  1.05    &  0.205(2) &  0.034(2) & (1, 8) &  0.183  \\
  32 &  1.19    &  0.200(1) &  0.049(2) & (1, 6) &  0.478  \\
  28 &  1.36    &  0.191(2) &  0.067(4) & (2, 6) &  0.297  \\
  24 &  1.58    &  0.189(2) &  0.083(5) & (2, 4) &  0.253  \\
  20 &  1.90    &  0.196(2) &  0.091(5) & (2, 4) &  0.046
\end{tabular}
\end{ruledtabular}
\end{table}

\begin{table}[tb]
\caption{The best-fit $\omega_0$ and $\Gamma$ of $T_2^{++}$ channel
at different $T$ through the Breit-Wigner fit. Also listed are the
fit window $[t_1, t_2]$ and $\chi^2/d.o.f$. \label{bw_opt17}}
\begin{ruledtabular}
\begin{tabular}{cccccc}
  $N_t$ & $T/T_{c}$ & $\omega_{0}$ & $\Gamma$ & $[t_1,t_2]$ & $\chi^{2}/d.o.f$  \\
  \hline
 128 &  0.30   &  0.210(1) &  0.008(2) & (3,10) &  0.442  \\
  80 &  0.47   &  0.213(1) &  0.009(3) & (3, 9) &  0.696  \\
  60 &  0.63   &  0.213(1) &  0.012(3) & (3, 7) &  0.326  \\
  48 &  0.79   &  0.210(1) &  0.007(4) & (3, 6) &  0.288  \\
  44 &  0.86   &  0.214(1) &  0.012(2) & (2, 6) &  0.437  \\
  40 &  0.95   &  0.208(1) &  0.023(1) & (1, 7) &  0.743  \\
&&&&&\\
  36 &  1.05    &  0.204(1) &  0.039(2) & (1, 7) &  0.606  \\
  32 &  1.19    &  0.199(1) &  0.047(1) & (1, 6) &  0.527  \\
  28 &  1.36    &  0.196(2) &  0.064(3) & (2, 5) &  0.022  \\
  24 &  1.58    &  0.194(1) &  0.077(4) & (2, 7) &  0.119  \\
  20 &  1.90    &  0.196(2) &  0.093(4) & (2, 5) &  0.027
\end{tabular}
\end{ruledtabular}
\end{table}

After the fit ranges for all the thermal correlators are chosen, the
jackknife analysis can be carried out straightforward and the
detailed procedures are omitted here. Table~\ref{bw_opt01},
~\ref{bw_opt03}, ~\ref{bw_opt09}, and ~\ref{bw_opt17} show the fit
windows $[t_1,t_2]$, the chi-square per degree of freedom
$\chi^2/d.o.f$, and the best-fit results of $\omega_0$ and $\Gamma$
at various temperature in $A_1^{++}$, $A_1^{-+}$, $E^{++}$, and
$T_2^{++}$ channels. In almost all the cases, the fit ranges start
from $t_1=1$, 2, or 3, and last for quite a few time slices. This
reflects that, as is expected, the optimal glueball operators couple
almost exclusively to the lowest spectral components after the
implementation of the variational method. All the $\chi^2/d.o.f$'s
are $\sim O(1)$ or even smaller, which reflect the reliability of
the fits.

The main features of the best fit $\omega_0$ and $\Gamma$ based on
Breit-Wigner $\emph{Ansatz}$ are described as follows:
\begin{itemize}
    \item The peak positions $\omega_0$ of the spectral
    functions $\rho(\omega)$ are insensitive to the temperature in
    all the considered channels. In particular, the $\omega_0$ in
    $A_1^{++}$ channel keeps almost constant all over the
    temperature range from $0.30T_c$ to $1.90T_c$. In the other
    three channels, the $\omega_0$'s do not change within errors
    below $T_c$, but reduce mildly with the increasing temperature
    above $T_c$. The reduction of $\omega_0$ at the highest
    temperature $T=1.90T_c$ is less than 5\% in these three
    channels.
    \item In all four channels,
    the thermal widths $\Gamma$ are small and do not vary much below $T_c$,
    but grow rapidly with the increasing temperature when $T>T_c$.
    Below $T_c$, the thermal widths are of order $\Gamma\sim 5\%$
    or even smaller (especially for the $A_1^{++}$ $\Gamma$ is
    consistent with zero). The thermal widths increase abruptly
    when the temperature passes $T_c$ and reach values $\sim
    \omega_0/2$ at $T=1.90T_c$.
\end{itemize}

These features can be seen easily in Fig.~\ref{opt_total_om} and
~\ref{opt_total_gm}, where the behaviors of $\omega_0$ and $\Gamma$
with respect to the temperature $T$ are plotted for all four
channels. The line$\_$shapes of the spectral functions with the
best-fit parameters at different $T$ are shown in Fig.~\ref{peak}
for $A_1^{-+}$, $E^{++}$, and $T_2^{++}$ channels (we do not plot
the spectral function of $A_1^{++}$ channel due to the small thermal
widths).

\section{Summary and Discussions}
\begin{table*}[ht]
\caption{The "pole masses" $M_G$ obtained by single-cosh analysis
and ($\omega_0, \Gamma$) obtained based on the Breit-Wigner
$\emph{ansatz}$ are combined together for comparison. Listed in the
table are the results in $A_1^{++}$ and $A_1^{-+}$ channels (all the
data are converted into the physical units).\label{comparison1}}
\begin{ruledtabular}
\begin{tabular}{cccccc|ccc}
       &         &           & $A_1^{++}$ &           & &       &  $A_1^{-+}$ &            \\
\hline
 $N_t$ & $T/T_c$ & $m_G$[GeV]& $\omega_0$[GeV] & $\Gamma$[GeV]& &   $m_G$[GeV]& $\omega_0$[GeV] & $\Gamma$[GeV]\\
\hline
 128   & 0.30     & 1.576(22) &    1.602(14)    &  0.091(113)&
                  & 2.488(31) &    2.549(17)    &  0.069(52)   \\
  80   & 0.47     & 1.621(29) &    1.644(22)    &  0.145(100)&
                  & 2.533(24) &    2.560(17)    &  0.086(44)   \\
  60   & 0.63     & 1.627(18) &    1.621(16)    &  0.098(38)&
                  & 2.499(27) &    2.559(16)    &  0.147(54)   \\
  48   & 0.79     & 1.616(21) &    1.612(26)    &  0.156(67)&
                  & 2.533(23) &    2.564(20)    &  0.135(55)   \\
  44   & 0.86     & 1.577(25) &    1.598(17)    &  0.045(34)&
                  & 2.454(34) &    2.577(18)    &  0.144(48)   \\
  40   & 0.95     & 1.576(39) &    1.621(20)    &  0.034(46)&
                  & 2.499(25) &    2.525(26)    &  0.042(71)   \\
 \hline
  36   & 1.05     & 1.486(43) &    1.638(28)    &  0.315(48)&
                  & 2.060(48) &    2.490(23)    &  0.413(32)   \\
  32   & 1.19     & 1.418(21) &    1.588(25)    &  0.586(28)&
                  & 1.959(37) &    2.464(20)    &  0.529(28)   \\
  28   & 1.36     & 1.373(48) &    1.599(26)    &  0.619(43)&
                  & 1.745(43) &    2.375(28)    &  0.768(71)   \\
  24   & 1.58     & 1.306(41) &    1.613(28)    &  0.664(37)&
                  & 1.644(45) &    2.308(32)    &  0.999(83)   \\
  20   & 1.90     &   \--     &    1.642(32)    &  0.873(59)&
                  &   \--     &    2.380(35)    &  1.114(99)
\end{tabular}
\end{ruledtabular}
\end{table*}

\begin{table*}[ht]
\caption{The pole masses $M_G$ obtained by single-cosh analysis and
($\omega_0, \Gamma$) obtained based on the Breit-Wigner
$\emph{Ansatz}$ are combined together for comparison.  Listed in the
table are the results in $E^{++}$ and $T_2^{++}$ channels (all the
data are converted into physical units). \label{comparison2}}
\begin{ruledtabular}
\begin{tabular}{cccccc|ccc}
       &         &           & $E^{++}$ &            &  &       &  $T_2^{++}$ &            \\
\hline
 $N_t$ & $T/T_c$ & $m_G$[GeV]& $\omega_0$[GeV] & $\Gamma$[GeV] &  & $m_G$[GeV]& $\omega_0$[GeV] & $\Gamma$[GeV]\\
 \hline
 128   & 0.30     & 2.364(11) &    2.385(12)    &  0.140(42)&
                  & 2.308(14) &    2.363(10)    &  0.091(25)   \\
  80   & 0.47     & 2.308(13) &    2.368(12)    &  0.069(29)&
                  & 2.353(15) &    2.387(10)    &  0.105(32)   \\
  60   & 0.63     & 2.330(11) &    2.383(12)    &  0.116(32)&
                  & 2.319(13) &    2.396(10)    &  0.140(32)   \\
  48   & 0.79     & 2.353(19) &    2.393(15)    &  0.129(34)&
                  & 2.308(15) &    2.362(14)    &  0.083(43)   \\
  44   & 0.86     & 2.319(15) &    2.379(12)    &  0.119(32)&
                  & 2.330(21) &    2.405(10)    &  0.136(26)   \\
  40   & 0.95     & 2.263(14) &    2.327(15)    &  0.247(38)&
                  & 2.218(16) &    2.344(11)    &  0.259(16)   \\
  \hline
  36   & 1.05     & 1.880(41) &    2.305(17)    &  0.382(23)&
                  & 1.925(33) &    2.298(14)    &  0.437(17)   \\
  32   & 1.19     & 1.722(35) &    2.247(16)    &  0.549(20)&
                  & 1.801(25) &    2.244(11)    &  0.532(15)   \\
  28   & 1.36     & 1.610(31) &    2.155(19)    &  0.754(43)&
                  & 1.666(23) &    2.205(17)    &  0.717(36)   \\
  24   & 1.58     & 1.565(23) &    2.132(21)    &  0.937(55)&
                  & 1.610(27) &    2.184(16)    &  0.870(41)   \\
  20   & 1.90     &   \--     &    2.201(23)    &  1.023(60)&
                  &   \--     &    2.209(20)    &  0.935(48)
\end{tabular}
\end{ruledtabular}
\end{table*}
On $24^3\times N_t$ anisotropic lattices with the anisotropy $\xi=5$
at the gauge coupling $\beta=3.2$, the thermal glueball correlators
are calculated in a large temperature range from $0.30T_c$ to
$1.90T_c$, which are realized by varying $N_t$ to represent
different temperatures. Based on the lattice spacing
$a_s=0.0878(4)\,{\rm fm}$ determined by $r_0^{-1}=(410(20)\,{\rm
MeV})$, the spatial extension of the lattices are estimated to be
$(2.1\,{\rm fm})^3$, which is large enough to be free of the finite
volume effects. On the other hand, because of the large anisotropy,
there are enough data point in the temporal direction for the
thermal correlators to be analyzed comfortably even at the highest
temperature $T\sim 2T_c$ concerned in this work. With the
implementation of the smearing scheme and the variational method, we
can construct the optimal glueball operators in all the symmetry
channel, which couple mostly to the lowest-lying states (or more
precisely, the lowest-lying spectral components). As a result, the
thermal correlators of these operators can be considered to be
contributed dominantly from these lowest-lying states. The thermal
correlators are analyzed based on two $\emph{ansatz}$, say, the
single-cosh function form and the Breit-Wigner $\emph{Ansatz}$. In
Table~\ref{comparison1} and Table~\ref{comparison2}, the "pole
masses" $M_G$ obtained by single-cosh analysis and ($\omega_0,
\Gamma$) obtained based on the Breit-Wigner $\emph{Ansatz}$ are
combined together for comparison (all the data are converted into
physical units).

The most striking observation from the single-cosh analysis is that,
in all 20 $R^{PC}$ channels, the best-fit pole-masses $M_G$ are
almost constant within errors from the low temperature up to right
below the critical temperature $T_c$. This is what should be from
the point of view of deconfinement phase transition of QCD: Since
below $T_c$ the system is in the confinement phase, the fundamental
degrees of freedom must be hadrons. Above $T_{c}$, the reduction of
the pole masses does signal the QCD transition, after which the
state of the matter is very different from that below $T_c$.
However, the existence of effective mass plateaus, from which the
pole masses are extracted, also implies that color singlet objects,
the glueball-like modes, can also survive at the intermediate
temperature above $T_c$. The results of the Breit-Wigner fit are
consistent with this picture. In the Breit-Wigner $\emph{Ansatz}$,
thermal widths $\Gamma$ are introduced to glueball states to account
for the effects of finite temperature, such as the thermal
scattering and the thermal fluctuations. As shown in
Table~\ref{comparison1} and~\ref{comparison2}, below $T_c$ (or in
the confinement phase), the best-fit $\omega_0$'s are very close to
the pole masses, and the thermal widths $\Gamma$ are very tiny and
are always of a few percent of $\omega_0$. This means the glueball
states are surely stable in the confinement phase and the thermal
interaction among them are weak. With the temperature increasing
above $T_c$, while the temperature dependence of $\omega_0$'s is
very mild, the thermal widths $\Gamma$ grow rapidly and reach values
of roughly half of $\omega$'s at $T\sim 1.9T_c$. This clearly
reflects that glueballs act as resonances are unstable more and
more, and the reduction of pole masses above $T_c$ can be taken as
the effect of these growing thermal widths.

To summarize, in pure gauge theory, the state of matter is dominated
by weakly interacting hadronlike states below $T_c$; when $T>T_c$,
glueball states survive as resonancelike modes up to a temperature
$T\sim 1.9T_c$ with their thermal widths growing with increasing
$T$, which implies that in this intermediate temperature range,
glueballs are unstable and may decay into gluons, and reversely
gluons also interact strongly enough to form glueball-like
resonances. The two procedure may reach the thermal equilibrium at a
given temperature, such that the gluon degree of freedom become more
and more important with $T$ increasing. At very high temperature,
the glueball-like resonances may disappear finally and the state of
matter can thereby be described by a perturbative gluon plasma. This
picture is coincident with the observations both in the study of
equation of state of QCD and the thermal properties of heavy
quarkonia. On the other hand, the surprising results of RHIC
experiments may also support this picture to some extent. First, the
data of RHIC experiments are well described by the hydrodynamical
model\cite{prl86}. Secondly, the investigation of elliptic flow data
using a Boltzmann-type equation for gluon scattering is not
consistent with the perturbative QCD apparently\cite{npa697}. So the
quark-gluon plasma at the RHIC temperature is most likely a strongly
interacting system.

\section*{Acknowledgments}
This work is supported in part by NSFC (Grant No. 10347110,
10421003, 10575107, 10675005, 10675101, 10721063, and 10835002) and
CAS (Grant No. KJCX3-SYW-N2 and KJCX2-YW-N29). The numerical
calculations were performed on DeepComp 6800 supercomputer of the
Supercomputing Center of Chinese Academy of Sciences, Dawning 4000A
supercomputer of Shanghai Supercomputing Center, and NKstar2
Supercomputer of Nankai University.

\end{document}